\documentclass[aps,twocolumn,prb,showpacs]{revtex4}
\usepackage{graphicx}
\usepackage{color}
\usepackage{amsmath}
\usepackage[bf,Large]{}

\begin{document}

\title{Spectral decomposition and matrix element effects in scanning 
tunneling spectroscopy of Bi$_2$Sr$_2$CaCu$_2$O$_{8+\delta}$.}

\author{Jouko Nieminen}
\email{jouko.nieminen@tut.fi}
\affiliation{Department of
Physics, Tampere University of Technology, P.O. Box 692, FIN-33101
Tampere, Finland}
\affiliation{Department of Physics, Northeastern
University, Boston}

\author{Ilpo Suominen}
\affiliation{Department of
Physics, Tampere University of Technology, P.O. Box 692, FIN-33101
Tampere, Finland}

\author{R.S. Markiewicz}
\affiliation{Department of Physics,
Northeastern University, Boston;
SMC-INFM-CNR, Dipartimento di Fisica, Universit\`a di Roma
 ``La Sapienza'', P. Aldo Moro 2, 00185 Roma, Italy;
ISC-CNR, Via dei Taurini 19, 00185 Roma}

\author{Hsin Lin} \author{A. Bansil}
\affiliation{Department of Physics,
Northeastern University, Boston}

\date{Version of \today}

\begin{abstract}

We present a Green's function based framework for modeling the scanning 
tunneling spectrum from the normal as well as the superconducting 
state of complex materials where the nature of the tunneling process$-$ 
i.e. the effect of the tunneling 'matrix element', is properly taken into 
account. The formalism is applied to the case of optimally doped 
Bi$_2$Sr$_2$CaCu$_2$O$_{8+\delta}$ (Bi2212) high-Tc superconductor using a 
large tight-binding basis set of electron and hole orbitals. The results 
show clearly that the spectrum is modified strongly by the effects of 
the tunneling matrix element and that it is not a simple replica of the 
local density of states (LDOS) of the Cu-$d_{x^2-y^2}$ orbitals with other 
orbitals playing a key role in shaping the spectra. We show how the 
spectrum can be decomposed usefully in terms of tunneling 'channels' or 
paths through which the current flows from various orbitals in the system 
to the scanning tip. Such an analysis reveals symmetry forbidden and 
symmetry enhanced paths between the tip and the cuprate layers. 
Significant contributions arise from not only the CuO$_2$ layer closest to 
the tip, but also from the second CuO$_2$ layer. The spectrum also 
contains a longer range background reflecting the non-local nature of the 
underlying Bloch states. In the superconducting state, coherence peaks are 
found to be dominated by the anomalous components of Green's function.

\end{abstract}

\date{Version of \today}

\maketitle

\section{Introduction}

High resolution scanning tunneling spectroscopy (STS) together with other 
highly resolved spectroscopies such as angle resolved photoemission 
(ARPES), is making it possible to obtain a comprehensive mapping of the 
electronic spectrum of the high-temperature superconductors (HTSs) in both 
real and reciprocal space over a wide range of dopings and temperatures. 
These studies are providing insight into the rich phase diagrams of the 
HTSs, and are leading thus to an understanding of the 'missing links' for 
developing a definitive theory of how high superconducting transition 
temperatures arise in these unconventional materials. In STS experiments, 
the focus to date has been on hole doped cuprates, especially on 
Bi$_2$Sr$_2$CaCu$_2$O$_{8+\delta}$ (Bi2212), which has been the subject of an 
overwhelming amount of experimental work, see, e.g, Refs. 
\onlinecite{Fischer,McElroy,Hudson,Pan,Yazdani,Balatsky1}.  Bi2212 is a 
typical cuprate material, which is an antiferromagetic insulator in the 
strongly underdoped regime, but exhibits a superconducting phase over a 
wide range of hole doping.

STS can be applied to a substantial part of the doping and temperature 
spanned phase space of HTS materials. The superconducting (SC) phase is 
observed around optimal hole doping (OP), while the pseudogap (PG) phase 
is found within the underdoped regime (UD).  As a practical limitation, 
STS requires a conducting sample, but the deeply underdoped regime is 
insulating and hence unreachable by STS. However, under experimental 
conditions the samples are not homogeneously doped. Rather, there is a 
strong spatial variation in doping, which makes observation of a continuum 
from the PG to the SC phase possible within one sample.  Although these 
spatial variations in STS generally appear irregular, quite recently a 
more ordered coexistence of PG and SC phases has been observed 
\cite{Kohsaka}.

The physics of the cuprates is dominated by the cuprate layers, which are 
usually not exposed to the tip of the apparatus. For example, in Bi2212, 
the quasiparticle tunneling takes place through insulating BiO and SrO 
layers.  The conventional interpretation of the spectra is based on the 
assumption that the STS spectrum is directly proportional to the LDOS of 
the CuO$_ 2$ layer, especially the LDOS of the $d_{x^2-y^2}$ orbitals, thus 
neglecting the effects of the tunneling process in modifying the spectrum 
in the presence of the insulating overlayers and multiple orbitals. The 
motivation for this simplification is an attempt to reduce the 
quasiparticle structure to few band models, which are amenable to 
theoretical treatment of strong correlation effects in the presence of 
superconducting and antiferromagnetic order. Notably, there have been 
attempts to take the effect of the overlayers into account by assuming a 
`tunneling matrix element' or a `filter function' \cite{Balatsky, 
hoogenboom,Fischer}. 

With this background, our recent work on STS\cite{NLMB} of Bi2212 provides 
a significant advance in realistic material-specific modeling of the STS 
spectrum. We invoke a Green's function approach where a {\it large} number 
of orbitals is included, and all tunneling paths to the tip in the 
semi-infinite solid are taken into account. We showed clearly that instead 
of being a simple reflection of LDOS of the Cu-$d_{x^2-y^2}$ orbitals, the STS 
signal represents a very complex mapping of the electronic structure of 
the system.

In this study we extend our approach by decomposing the tunneling
current in terms of regular and anomalous matrix elements of the
spectral function in an atomic orbital basis. As in Ref.
\onlinecite{NLMB}, we concentrate on Bi2212 as the canonical HTS
material. We start by reformulating the well-established methods to
model tunneling current in nanostructures into a more transparent form
for interpreting tunneling in the superconducting state. Our
derivation is based on the conventional Todorov-Pendry
\cite{Todorov,Pendry} approach (TP), which is closely related to the
more common Tersoff-Hamann \cite{Tersoff} method (TH). TP and TH
methods both employ a calculation of the LDOS, but TP is more
naturally written in terms of Green's functions.  We will show, in
fact, that TP decomposes into matrix elements of the spectral
function, giving very detailed information concerning the origin of
various features in the tunneling spectrum. 
We thus demonstrate how the contribution of different
atomic orbitals to the total current can be extracted from the
calculations. Our spectral decomposition also naturally distinguishes
between the electron and hole nature of the quasiparticles in the
superconducting state. In addition, it leads to a
multiband generalization of filtering function by Martin et al
\cite{Balatsky} and a clarification of selection rules
governing tunneling through filtering layers.  This information is
important, e.g., in determining how a dopant or impurity atom alters
the spectrum, and how the effect of such a perturbation is seen in
real space.

In order to gain a handle on the effects of filtering layers, we derive a 
consistent form of a filter function through Green's function 
manipulations. This rigorous form for the filtering effects is useful for 
determining the relation between the tunnel current and the LDOS of the 
CuO$_2$ layers.  We show that this relation is nontrivial in that some 
channels are `first-order forbidden'.  Thus our new approach shows that no 
direct regular signal from $d_{x^2-y^2}$ orbitals of the Cu directly below 
the STM tip reaches the microscope. Instead the $d_{x^2-y^2}$ orbitals of 
the four neighboring Cu atoms give a major contribution to the tunneling 
signal.  Although we concentrate on pristine systems in the present work, 
the results have important implications for inhomogeneous situations -- 
e.g., the relationship between the observed features in the spectrum of an 
impurity atom and the underlying LDOS.  This decomposition also allows 
treatment of the regular and anomalous propagation of quasiparticles in a 
superconductor, and on this basis we show that the coherence peaks result 
from the anomalous electron-hole propagation.

The paper is organized as follows. The model for the geometrical
structure and the electronic structure is introduced in Sections~II.A
and B, respectively. The methods to calculate the Green's function in
the normal and the superconducting state are derived in Sections II.B
and C, respectively.  The Todorov-Pendry equation for the tunneling
current is decomposed into regular and anomalous terms to show not
only the proper form of the matrix element but also the partial
current terms for any chosen orbital in Section II.D. The formalism is
applied to discuss STM topographic maps in Section~III.A, and the STS
spectrum of Bi2212 in Section~III.B. The spectrum is then analyzed in
terms of tunneling matrix elements and partial currents in
Section~III.C. Further comments on symmetry analysis are made in
Section~IV.A., and remarks on electron extraction/injection are made
in Section~IV.B. Finally, conclusions are drawn and future
applications sketched in Section~V. Relevant technical details of the
form of boson-electron coupling assumed in the tunnel spectra and of
the superconducting state calculations are given in the two
appendices.

\begin{figure}[th]
\centering \includegraphics[width=0.5\textwidth]{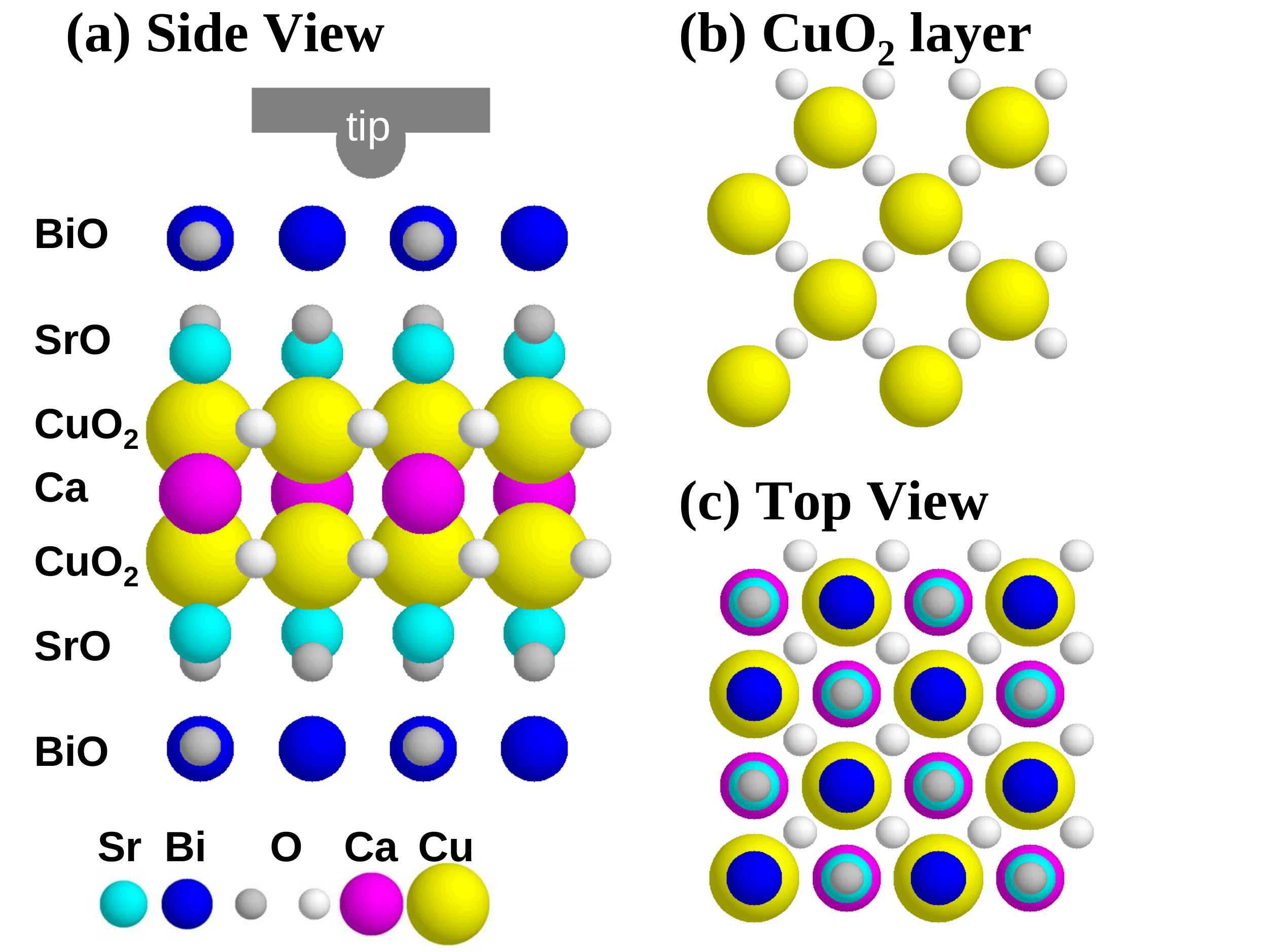}
\caption{(color online) 
(a) Side view of the simulation cell used to compute the tunneling 
spectrum of Bi2212. Tunneling signal from the conducting CuO$_2$ layers 
reaches the tip after passing through the filtering layers of SrO and BiO. 
(b) Cuprate layer showing the supercell consisting of eight primitive 
cells. (c) Top view of the surface showing the arrangement of various 
atoms.}
\label{geometric}
\end{figure}

\section{Description of the model}

Our theoretical framework involves three distinct steps. First, we choose 
a three-dimensional geometrical model of atoms with a sufficiently large 
simulation cell with periodic boundary conditions in the horizontal 
directions to treat a semi-infinite solid surface. Second, we attach a 
basis set of atomic orbitals to each atom. At this stage, the one-particle 
Hamiltonian is constructed and the corresponding Green's function tensor 
is formed. Third, we apply our Green's function formalism to evaluate 
the tunneling current. The technical details of these three steps are 
outlined in the following three subsections.

\subsection{Sample geometry}

We model the Bi2212 sample as a slab of seven layers \cite{footnote1}
in which the topmost
layer is BiO, followed by layers of SrO, CuO$_2$, Ca, CuO$_2$, SrO,
and BiO, as shown in Fig.  \ref{geometric}(a). The tunneling
computations are based on a $2\sqrt{2} \times 2\sqrt{2}$ real space
supercell consisting of 8 primitive surface cells with a total of 120
atoms (see Fig. \ref{geometric}(b)). The coordinates are taken from
the tetragonal crystal structure of Ref.  \onlinecite{Bellini}. For
STS simulations, the STM tip is modeled as an orbital with an s-wave
symmetry at the assumed position of the apex of the tip.  This tip is
allowed to scan across the substrate for generating the topographic
maps such as those in Fig. \ref{exptheo}, or held fixed on top of a
surface Bi atom for the computed spectra presented for example in
Fig. \ref{pristine}.


\subsection{Construction of the uncorrelated normal state Hamiltonian}

In order to construct a realistic framework capable of describing the
tunneling spectrum of the normal as well as the superconducting state
of the cuprates, we start with the normal state Hamiltonian for the
semi-infinite solid in the form
\begin{equation}
\hat{H}_1 = \sum_{\alpha\beta\sigma}
\left[\varepsilon_{\alpha}c^{\dagger}_{\alpha \sigma} c_{\alpha \sigma}+
V_{\alpha \beta}
c^{\dagger}_{\alpha \sigma} c_{\beta\sigma}\right],\label{H1}
\end{equation}
which describes a system of tight-binding orbitals created (or
annihilated) via the real-space operators $c^{\dagger}_{\alpha
\sigma}$ (or $c_{\alpha \sigma}$). Here $\alpha$ is a composite index
denoting both the type of orbital (e.g. Cu-$d_{x^2-y^2}$) and the site
on which this orbital is placed, and $\sigma$ is the spin index.
$\varepsilon_\alpha$ is the on-site energy of the $\alpha^{th}$
orbital. $\alpha$ and $\beta$ orbitals interact with each
other through the potential $V_{\alpha\beta}$ to create the energy
eigenstates of the entire system.

\begin{figure}[th]
 \centering \includegraphics[width=0.5\textwidth]{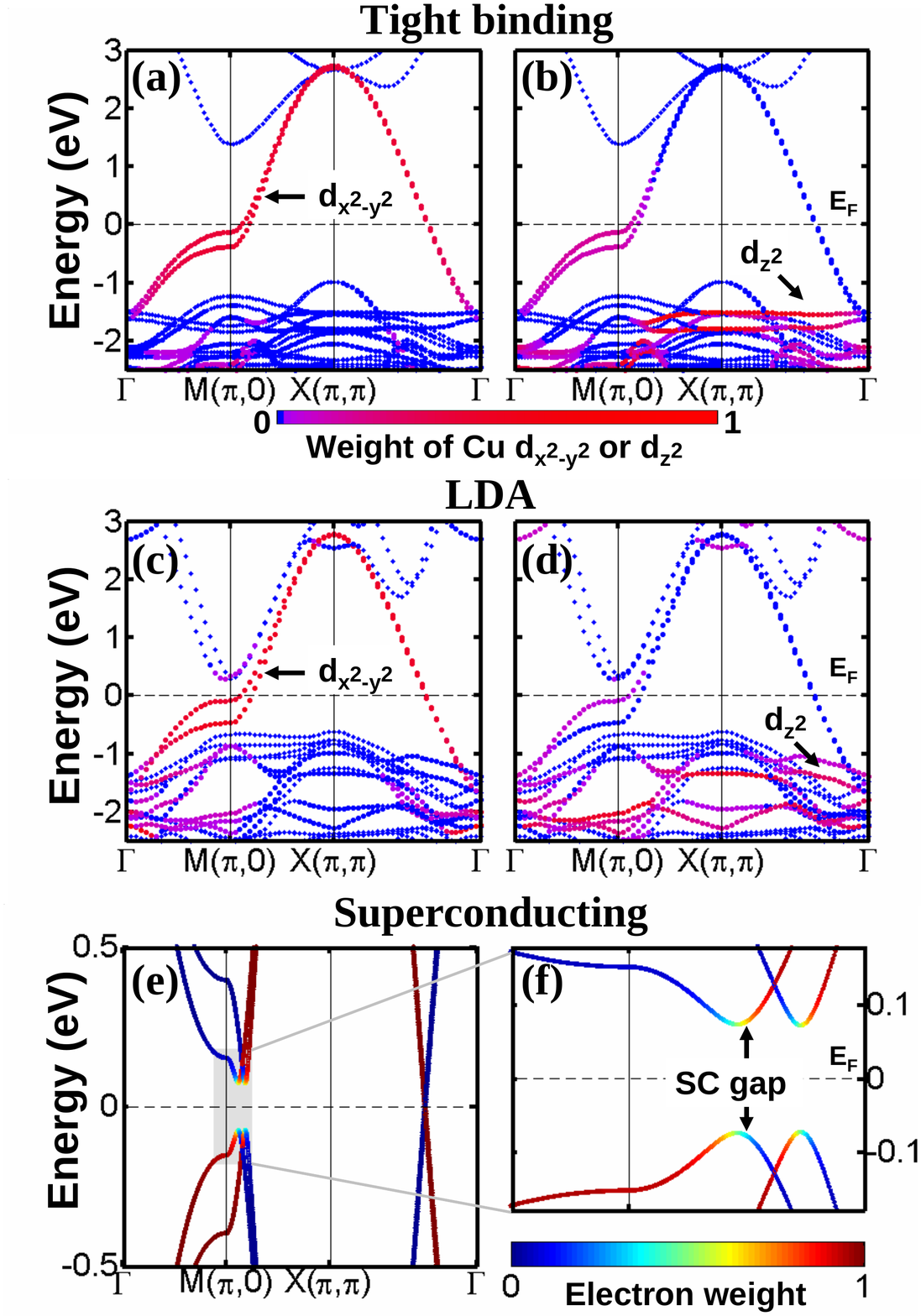}
\caption{(color online) (a)-(d): Normal state band structure of Bi2212
for the tight-binding Hamiltonian and from first-principles LDA
computations. Weights of Cu-$d_{x^2-y^2}$ and Cu-$d_{z^2}$
contribution to the bands are shown using a colorscale where red
denotes high and blue low values (see colorbar). Note that the
tight-binding calculations are done for a slab, so that the the
tight-binding bands do not display the splitting of Bi-O bands seen in
LDA results. The LDA bands have been calculated using Virtual Crystal
Approximation (VCA) with 24\% Pb doping to set the bottom of the BiO
band. (e) and (f): Quasiparticle band structure in the superconducting
state based on the Hamiltonian of Eq. \eqref{hamiltonian} is shown in
(e). Panel (f) zooms in on the gap region of (e) which is shaded grey.
Electron character of quasiparticles is shown in red and the hole
character in blue.  Notice, that the quasiparticles differ
significantly from being electrons or holes only in the close
neighborhood of the superconducting gap around the M-point.  }
\label{norbands}
\end{figure}

The specific electron and hole orbital sets used for various atoms
are: ($s,p_x,p_y,p_z$) for Bi, Ca and O; $s$ for Sr; and
($4s,d_{3z^2-r^2},d_{xy},d_{xz},d_{yz}, d_{x^2-y^2}$) for Cu
atoms. This yields 58 electron or hole orbitals in a primitive cell
and a total of $2 \times 464$ orbitals in the $2\sqrt{2} \times
2\sqrt{2}$ simulation supercell. The number of {\bf k}-points used in
the computations depends on whether we do band calculations or solve
the Green's function. For band calculations, we use a dense set of
{\bf k}-values to produce smooth bands for directions $\Gamma
\rightarrow M \rightarrow X \rightarrow \Gamma$ as seen for example in
Fig. \ref{norbands}. In the case of Green's function calculations, we
use $N_k=256$ {\bf k}-points for the supercell Brillouin zone. This
corresponds to $8 \times 256 = 2048$ {\bf k}-points for a primitive
cell.

\begin{table}[th]
\begin{tabular}{|c|c|c|c|c|c|c|c|c|c|}
\hline
$v_{\alpha \beta m} (eV)$  &\multicolumn{8}{c}{} & \\
\hline
$v_{ss\sigma}$  &
$v_{sp\sigma}$  &
$v_{pp\sigma}$  &
$v_{pp\pi}$     &
$v_{sd\sigma}$  &
$v_{pd\sigma}$  &
$v_{pd\pi}$     &
$v_{dd\sigma}$  &
$v_{dd\pi}$     &
$v_{dd\delta}$  \\
\hline
 -0.28  &
 0.94   &
 1.23   &
-0.13   &
-0.62   &
-2.81   &
 1.16   &
-9.00   &
 12.60  &
-2.29   \\
\hline
\end{tabular}\\

\begin{tabular}{|c|c|c|c|c|c|c|}
\hline
$\varepsilon_{\alpha} (eV)$  &\multicolumn{5}{c}{}& \\
\hline
s/Bi&
p/Bi&
s/O(Bi)&
p/O(Bi)&
s/Sr&
s/Ca&
p/Ca\\
\hline
-12.200&  1.800&
14.700&  -2.400&
7.819&
5.631&  13.335a\\
\hline
s/O(Sr)&
p/O(Sr)&
s/Cu&
d/Cu&
s/O(Cu)&
p/O(Cu)&
\\
\hline
-15.270&  -2.353&
5.001&  -2.962&
-18.560&  -3.825&
\\
\hline
\end{tabular}\\

\caption{Slater-Koster prefactors, $v_{\alpha \beta m},$ and onsite energies
$\varepsilon_{\alpha}.$  The $v_{\alpha \beta m}$ are used to
construct the Hamiltonian overlap matrix elements $V_{\alpha\beta}$ as
described in Ref. \cite{Slater}.}
\end{table}

The Slater-Koster formalism \cite{Slater, Harrison, Shi} is used to
fix the angular dependence of the tight binding overlap integrals. The
onsite energies and the prefactors are fitted to the LDA band
structure of Bi2212 that underlies for example the extensive
angle-resolved photointensity computations of Refs.
~\onlinecite{bansil99,lindroos02,bansil05,markiewicz05,asensio03,bansil98}.
In Table I, we show the specific values of the $v_{\alpha \beta m}$
prefactors used for computing the Slater-Koster hopping integrals.
Notably, we have shifted the bottom of the BiO conduction band to
agree with experiments, which do not observe the Bi-bands at least
within $1eV$ above the Fermi-level. This choice is also supported by
calculations of Ref.  \onlinecite{HL}, which show the sensitivity of
the position of the Bi-band with respect to impurities and doping. The
absence of the bottom of the BiO band in the STS spectra may also be
due to a voltage gradient across the insulating filter layers (BiO and
SrO layers) when applying a bias voltage between the tip and the
sample. If so, the absolute value of the voltage within these layers
is less than the bias voltage $V_b$, and thus the apparatus would need
to apply a bias which would be significantly larger than $V_b$ to
locally see states that are strictly at $E_F+eV_b.$

The tight-binding parameters of the normal state Hamiltonian of Eq. 
\eqref{H1} produce the detailed band structure of Bi2212 shown in Fig. 
\ref{norbands}.  While the tight-binding band structure is in reasonable 
agreement with the LDA band structure of Ref.~\onlinecite{HL}, in order to 
carry out spectroscopic computations, one must additionally make sure that 
the underlying wavefunctions are described correctly including their 
symmetries. Our procedure based on the the use of Koster-Slater matrix 
elements not only fits the band stuctures, but the symmetries and 
phases of the associated wavefunctions are also described correctly.

Figs. \ref{norbands} (a)-(b) show the normal state tight-binding band
structure based on our $58$ orbital Hamiltonian of Eq. \eqref{H1}. The
main cuprate bands, with predominantly Cu-$d_{x^2-y^2}$ character, are
seen in panels (a) and (b) to follow the corresponding LDA
calculations in panels (c) and (d). Note that in our tight-binding
modeling, we have adjusted the positions and bilayer splitting of the
two van Hove singularities (VHSs) to approximately match the
experimental photoemission and STS findings for the optimal doping
(OP) region with hole concentration $p \approx 0.16$
(Ref. \onlinecite{Gomes, Kaminski}). In addition to Cu-$d_{x^2-y^2}$,
Cu-$d_{z^2}$ is seen in panels (b) and (d) to give a significant
spectral weight to this band, especially at energies below the
Fermi-level. The complicated `spaghetti' region has large
contributions from the $d_{z^2}$ of Cu and horizontal $p_x(p_y)$
orbitals of the oxygens within the cuprate layer as well as the
vertical $p_z$ orbital of the apical oxygen.  Concerning the filter
layers, the bottom of the BiO-like conduction band (or bismuth pocket) along
the $M(\pi ,0)$ direction carries the character of the horizontal p-orbitals of the
surface oxygens $O(Bi)$ (see Fig. ~\ref{norbands} (a)).

In tunneling calculations, we directly evaluate the
Green's function instead of diagonalizing the Hamiltonian.
For this purpose, the normal state Green's function is solved first by 
starting from the orbital matrix elements of the Green's function:
\begin{equation}
g^{\pm}_{\alpha \beta} = \frac{\delta_{\alpha \beta}}{\varepsilon -
\varepsilon_{\alpha}  - \Sigma^{\pm}_{\alpha}(\varepsilon)},
\label{independent}
\end{equation}
where $\varepsilon_{\alpha}$ is the onsite energy of the orbital
$\alpha.$ At this point, a diagonal self-energy
$\Sigma^{\pm}_{\alpha} =\Sigma_{\alpha}{'} \pm i \Sigma_{\alpha}{''}$ can 
be included straightforwardly. 
The simplest self-energy is a constant broadening of the
states in the form of a convergence factor $\Sigma^{\pm}_{\alpha} =
\mp i\eta.$  Appendix A (Eq. \eqref{sigmaeinstein}) presents a more general
self-energy which we use to model electron-boson coupling.

The total Green's function $G$ is constructed by solving Dyson's equation
\begin{displaymath}
G = g + gVG,
\label{Dyson}
\end{displaymath}
where $V_{\alpha \beta}$ are the off-diagonal overlap integrals of
Eq. \eqref{H1} Dyson's equation is exactly solved using the method described
in Ref. \cite{Nieminen}, which is suitable for tunneling calculations
\cite{footfourier}.

\subsection{Pairing interaction and the superconducting state 
Hamiltonian}

Superconductivity is included by adding a pairing interaction term
$\Delta$ in the Hamiltonian of Eq. \eqref{H1} as follows
\begin{equation}
\hat{H} = \hat{H}_1 + \sum_{\alpha \beta
\sigma} \left[\Delta_{\alpha \beta} c^{\dagger}_{\alpha \sigma}
c^{\dagger}_{\beta -\sigma} + \Delta_{\beta \alpha}^{\dagger}
 c_{\beta -\sigma} c_{\alpha \sigma} \right]
\label{hamiltonian}
\end{equation}
A gap parameter value of $\vert\Delta\vert = 0.045 eV$ is chosen to
model a typical experimental spectrum\cite{McElroy} for the illustrative 
purposes of this study.  We take $\Delta$ to be non-zero only between
$d_{x^2 - y^2}$ orbitals of the nearest neighbor Cu atoms, and to
possess a d-wave form, i.e., $\Delta_{d (d \pm x)} = +\vert \Delta \vert$
and $\Delta_{d (d \pm y)} = -\vert \Delta \vert,$ where $d$ denotes the
$d_{x^2-y^2}$ orbital at a chosen site, and $d \pm x/y$ the
$d_{x^2-y^2}$ orbital of the neighboring Cu atom in x/y-direction.
In momentum space, the corresponding $\Delta$ is given by
\begin{equation}
 \Delta_k =  \frac{\Delta}{2} \left[\cos{k_x a} - \cos{k_y a} \right],
\end{equation}
where $a$ is the in-plane lattice constant.
The pairing interaction of Eq.~\eqref{hamiltonian} allows electrons of
opposite spins to combine to produce superconducting pairs such that
the resulting superconducting gap is zero along the nodal directions
$k_x=\pm k_y$, and is maximum along the
antinodal directions. This choice of pairing interaction follows,
e.g., the one-band formalism given in Ref. \onlinecite{Flatte}.

\begin{figure}[h]
    \includegraphics[width=0.50\textwidth]{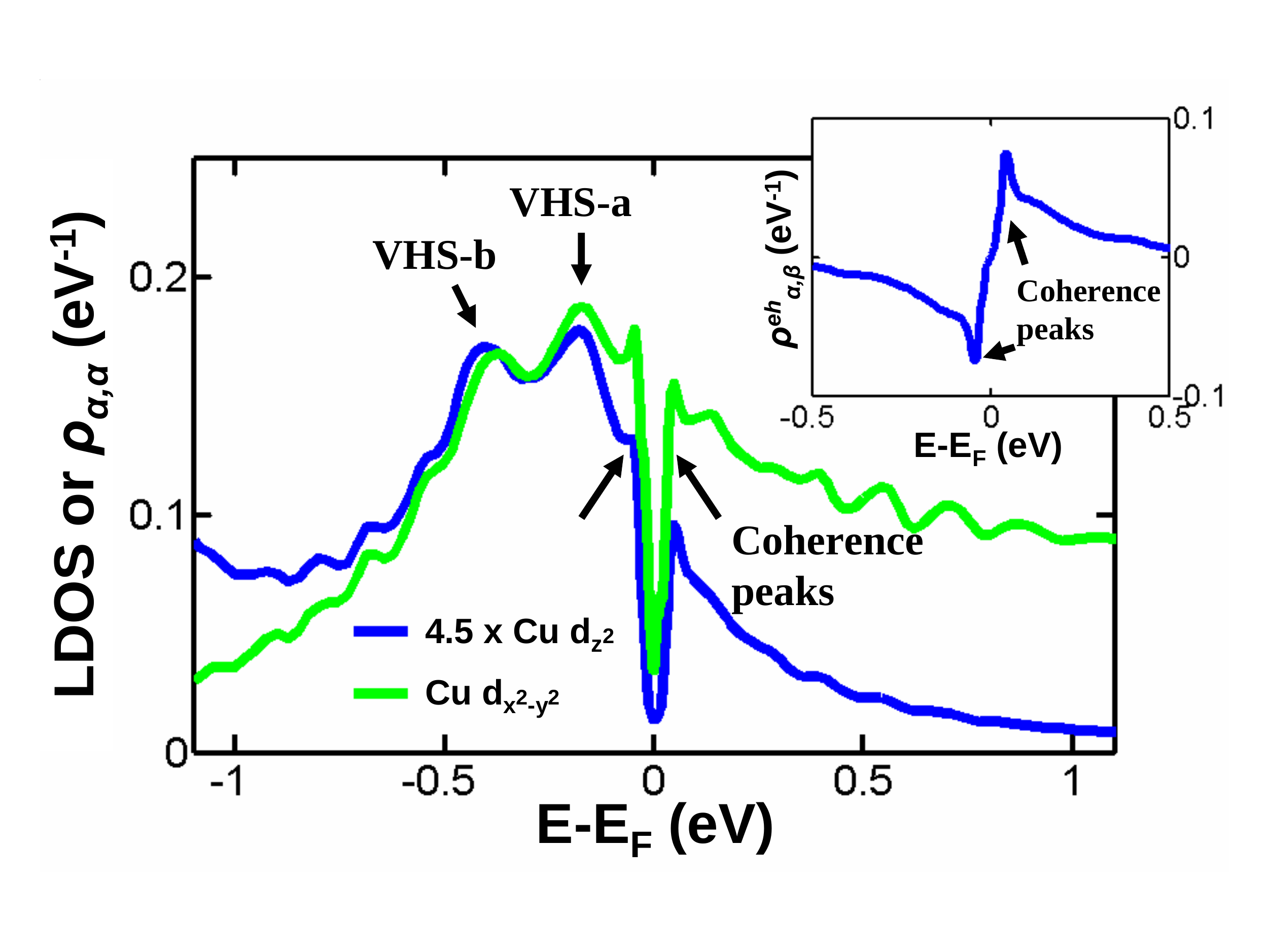}
  \caption{(color online) {\it Main:} LDOS (or the diagonal elements
$\rho_{\alpha \alpha}$ of the density matrix; see Appendix B for
details) of $d_{x^2-y^2}$ (green) and $d_{z^2}$ (blue) orbitals of
Cu. [Note $d_{z^2}$ curve is scaled up by a factor of 4.5 to compare
the shapes of the two LDOSs.] Oscillations at high positive or
negative energies ( above $\pm 0.5$ eV) are artifacts due to the use of
a sparse mesh of {\bf k}-points in the computation.  {\it Inset:}
Anomalous density matrix term $\rho^{eh}_{\alpha \beta}$ discussed in
Appendix B, where $\alpha$ and $\beta$ denote $d_{x^2-y^2}$ orbitals
of two neighboring Cu atoms.}
\label{ldos400403}
\end{figure}

For treating the superconducting case, we employ the tensor
(Nambu-Gorkov) Green's function ${\cal G}$ (see
Ref. \onlinecite{Fetter}) with the corresponding Dyson's equation:
\begin{equation}
  {\cal G} =  {\cal G}^0 +  {\cal G V G}^0,
\label{dyson1}
\end{equation}
where
\begin{displaymath}
  {\cal G} =
\left(
   \begin{array}{cc}
G_{e}& F\\
F^{\dagger}& G_{h}
   \end{array}
\right)~\textrm{and}~
{\cal V} =
\left(
  \begin{array}{cc}
0& \Delta\\
\Delta^{\dagger}& 0
  \end{array}
\right)
\end{displaymath}
where $G_{e}$ and  $G_{h},$ denote the Green's functions for the electrons 
and holes,
respectively.

The normal state electron Green function $G_{e}$ can be used to derive the
hole Green function $G_{h}$.
It can be shown by, e.g., the equation of motion method, that
\begin{displaymath}
G^{\pm}_{h,\alpha \beta}(E) = -G^{\mp}_{e, \beta \alpha}(-E)
\end{displaymath}
It is straightforwardly shown then that
  \begin{eqnarray}
   G_{e}& =& G_{e}^0 + F\Delta^{\dagger} G_{e}^0 \nonumber\\
F&= & G_{e} \Delta G_{h}^0
\label{fdelta}
\end{eqnarray}
The quasiparticle Green's function projected onto
electron degrees of freedom is then written in the form
\begin{equation}
  G_{e} = G_{e}^0 + G_{e} \Sigma^{BCS} G_{e}^0,
~\textrm{where}~
  \Sigma^{BCS} = \Delta G_{h}^0 \Delta^{\dagger}.
\label{bcsself}
\end{equation}

We also need the self-energy term $\Sigma^{h}_{\alpha}$ for
holes.  Since the transformation from electron to holes follows that
of the Green's function, we obtain the general form
\begin{displaymath}
  \Sigma^{h}_{\alpha}(\varepsilon) = -
\Sigma^{e*}_{\alpha}(-\varepsilon) =
-\Sigma^{'}_{\alpha}(-\varepsilon) + i
\Sigma^{''}_{\alpha}(-\varepsilon).
\end{displaymath}
In our particular case, we use a self-energy with an odd real part and
an even imaginary part as discussed in Appendix A 
(see Eq.  \eqref{sigmaeinstein})
Our self-energy is thus invariant under electron-hole
transformation.

\begin{figure}[h]
\centering
    \includegraphics[width=0.50\textwidth]{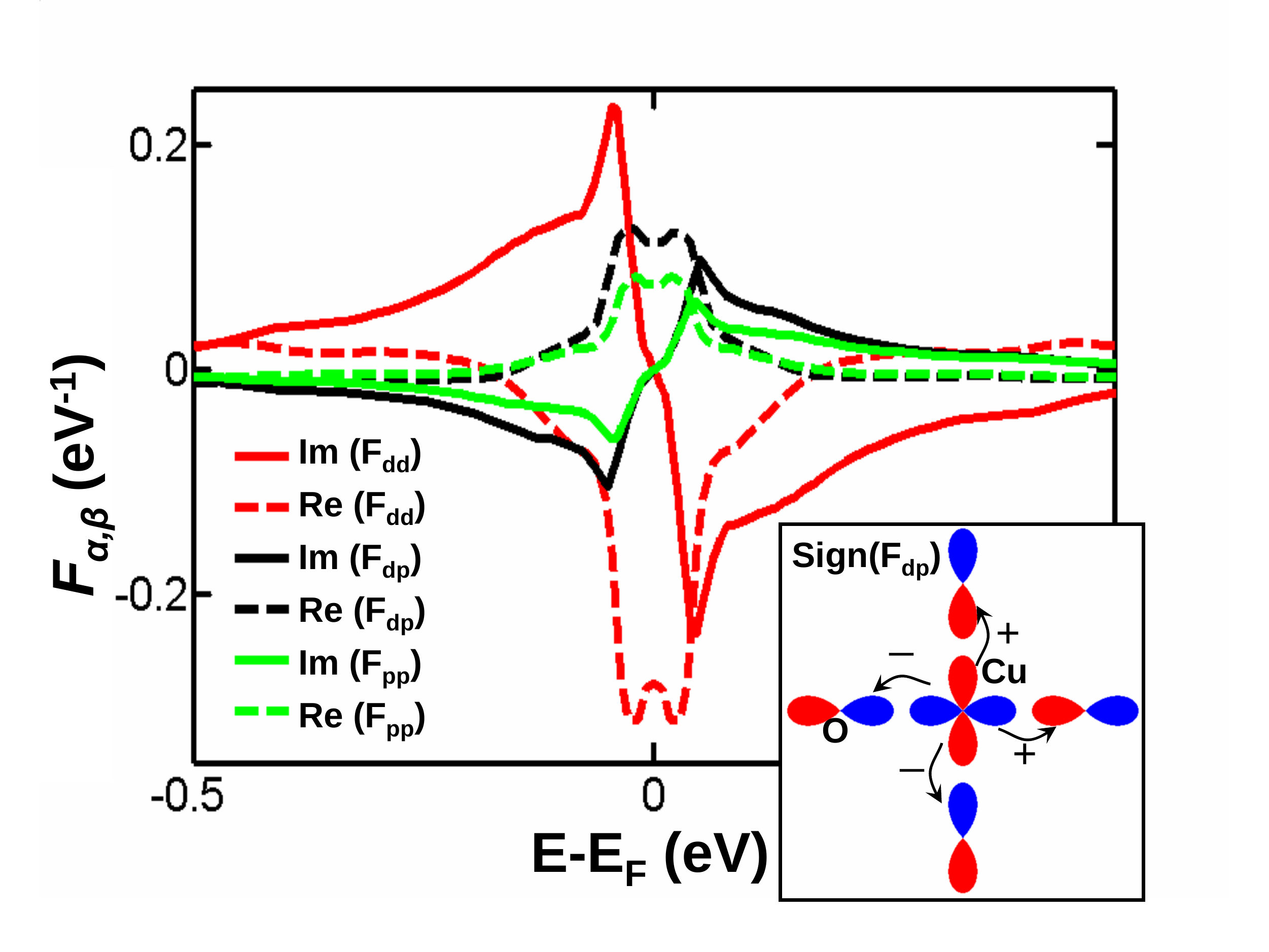}
  \caption{(color online) {\it Main:} Matrix elements of the anomalous
Green's function for onsite $p_x$-orbital of an intermediate oxygen
atom (green lines), $d_{x^2-y^2}$ orbitals of two neighboring Cu atoms
(red lines), and between Cu-$d_{x^2-y^2}$ and a $p_x$ orbital of a
neighboring oxygen (black lines). {\it Inset:} The directional
dependence of the sign of the off-diagonal element $F_{dp}$. For
details see special case (3) in the text.}
  \label{offdiag}
\end{figure}

Figs. \ref{norbands}(e) and (f) show the modifications of the normal
state band structure from the introduction of the pairing
interaction. Only the region within $\pm 500 meV$ of the Fermi level
is shown in panel (e), as the remainder of the bands are unchanged
from the normal state results of panels (a) and (b). The
superconducting state dispersion in panels (e) and (f) clearly
displays a d-wave gap with a maximum in the antinodal region near the
$M$ point and zero gap along the nodal direction near $(\pi /2,\pi
/2)$.  Note that both bonding and antibonding VHSs possess gaps of
similar magnitude. Fig. \ref{norbands}(e) also shows the relative
electron/hole character of the quasiparticles. As expected, the
quasiparticles are very distinctly either electron- or hole-like
almost everywhere except within a very narrow energy range at the top
and bottom of the SC gap. Fig. \ref{ldos400403} further shows that
mixing of the electron and hole features gives rise to coherence peaks
in the LDOS of Cu-$d_{x^2-y^2}$ and to a lesser extent in the LDOS of
Cu-$d_{z^2}$. The effects of electron-hole mixing are however most
pronouned in the anomalous matrix element of the quasiparticle Green's
function (inset to Fig. \ref{ldos400403} and Fig.~\ref{offdiag}). In
fact, the off-diagonal matrix element between an up-spin $d_{x^2-y^2}$
electron orbital and a down-spin $d_{x^2-y^2}$ hole orbital of two
neighboring Cu atoms gives the most important term in the anomalous
part of the Green's function. This term has d-wave symmetry, which
manifests itself as a change in sign each time we make a rotation of
$\frac{\pi}{2}$ around the central Cu site. In addition to the
coherence peaks, the anomalous density matrix inherits features from
the VHSs in the regular part of the density matrix, which in view of
electron-hole symmetry are reflected on both sides of the Fermi
energy. Additionally, strong hybridization between up-spin
Cu-$d_{x^2-y^2}$ electron orbitals and down-spin orbitals of O~$p_{x}$
holes (and vice versa) takes place as shown in
Fig. \ref{offdiag}. This term is comparable in strength to the
Cu-$d~-~$Cu-$d$ terms and changes sign in rotations of $\pi$ for reasons
explained in the special case (3) of the following paragraph. 
Fig.~\ref{offdiag} also shows a small onsite contribution
from the up-spin electron and down-spin hole of the $p_x$-orbital on
the oxygen between two neighboring Cu atoms. It is notable that these
matrix elements strictly follow the d-wave symmetry in rotations
around the central Cu atom.

These transformation properties follow consistently from Eq. 
\eqref{fdelta}.  Let us, for example, look at the equation in the 
$x$-direction: $F_{\alpha \beta} = G_{e,\alpha d}\Delta_{d (d\pm x)} 
G^{0}_{h,(d \pm x)\beta},$ where $d$ is a shorthand notation for 
$d_{x^2-y^2}$ of a chosen Cu atom, and $d \pm x$ stands for the 
$d_{x^2-y^2}$ orbital of the neighboring Cu atom in the 
positive/negative x-direction and consider several specific cases as 
follows.\\
\indent (1)
For $\alpha=d$ and $\beta=d\pm x$, both $G_{e,\alpha d}$
and $G^{0}_{h,(d \pm x)\beta}$ are onsite matrix elements, and thus
their sign remains invariant when changing from one Cu to another.
Hence the term $\Delta_{d(d\pm x)}$ is decisive, and the sign
can change only in going from x- to y- direction;\\
\indent (2) For $\alpha= \beta=$ O-$p_x,$ we have to first look at the term 
$G_{e,p_x d} G^{0}_{h,(d \pm x) p_x}$. Since the relative phases of the
off-diagonal matrix elements of the Green's function are proportional to
the sign of the overlap of the two orbitals, it is straightforward to
see from the signs of the lobes of the $d$ and $p$ orbitals that this
product is invariant to change in direction as well as in going from 
x to y. Therefore, $\Delta_{d (d\pm x)}$ again gives the d-wave symmetry
of these terms;\\
\indent (3) For $\alpha = d$ and $\beta =$ O-$p_{x}$,
$G_{e,\alpha d}$ is diagonal and thus invariant. Considering the
overlaps, one sees that 
$$G^{0}_{h,(d - x) \beta} = -G^{0}_{h,(d + x) \beta},$$ and 
$$G^{0}_{h,(d + y) \beta} = -G^{0}_{h,(d - y) \beta} =
-G^{0}_{h,(d + x) \beta}.$$
But, since $\Delta_{d (d\pm x)} = -\Delta_{d
(d\pm y)}$, 
$$F_{d p_{x}(-)} = -F_{d p_{x}(+)} = -F_{d
p_{y}(+)}= F_{d p_{y}(-)},$$ as shown in the inset to Fig.
\ref{offdiag}.

Eq. \eqref{ccF} of Appendix B shows that 
$F_{\alpha \beta}\propto\langle c_{\alpha\uparrow}c_{\beta \downarrow}
\rangle.$ Hence, case (3) of the last paragraph indicates that there
is a significant pairing $\langle c_{d_{x^2-y^2} \uparrow}c_{\phi
  \downarrow} \rangle$ when 
$$\vert \phi \rangle \propto \vert p_{x}(+) \rangle + \vert p_{y}(+)
\rangle - \vert p_{x}(-)\rangle - \vert p_{y}(-) \rangle.$$ Recall
that we introduced superconductivity in Hamiltonian of Eq. \eqref{hamiltonian}
{\it only} on the Cu-$d_{x^2-y^2}$ orbitals.  Thus we see that within our 
model the strong Cu-O hybridization automatically induces pairing on
the oxygen orbitals. This pairing is analogous to the concept 
of Zhang-Rice singlets (ZRS) in the low doping limit \cite{Zhang}, where 
pair states
$$
\vert d_{x^2-y^2} \uparrow \rangle \vert \phi \downarrow\rangle -
\vert d_{x^2-y^2} \downarrow \rangle \vert \phi \uparrow\rangle $$
are formed. Note, however, that ZRS is a concept related to
doping levels in the `normal' phase, and is not directly concerned
with superconductivity.  Nevertheless, the preceding 
considerations indicate that our model
is in accord with the ZRS scenario of the normal state \cite{onsitep}.

\subsection{Green's function formulation of tunneling current}

We turn now to consider the formulation of the
tunneling spectrum. For this purpose, we apply the conventional form of
the Todorov-Pendry expression \cite{Todorov,Pendry} for the
differential conductance $\sigma$ between orbitals of the tip ($t,t'$)
and the sample ($s,s'$), which in our case is straightforwardly shown
to yield
\begin{equation}
\sigma = \frac{dI}{dV} = \frac{2 \pi e^2 }{ \hbar} \sum_{t t' s s'}
\rho_{tt'}(E_F)V_{t's} \rho_{ss'}^{}(E_F+eV)V_{s't}^{\dagger},
\label{conductance}
\end{equation}
where the density matrix
\begin{equation}
\rho_{s s'} = -\frac{1}{\pi}Im[G_{s s'}^{+}]
 = \frac{1}{2\pi i} \left( G^{-}_{s s'}
- G^{+}_ {s s'} \right) ,
\label{spectralfunctiona}
\end{equation}
is given in terms of the retarded electron Green function or
propagator $G_{s s'}^{+}$. Eq.  \eqref{conductance} differs from the
more commonly used Tersoff-Hamann approach\cite{Tersoff} in that it
takes into account the details of the symmetry of the tip orbitals and
how these orbitals overlap with the surface orbitals.

Since electrons are not eigenparticles in the presence of the pairing 
term, Dyson's equation needs to be applied to the Green's function tensor:
\begin{equation}
    {\cal G}^{-} =   {\cal G}^{+} +
{\cal G}^{+}({\mathbf \Sigma^{-}}-{\mathbf\Sigma^{+}}){\cal G}^{-} =
{\cal G}^{+} - 2i{\cal G}^{+}{\mathbf \Sigma^{''}}{\cal G}^{-}
\label{spectaltensor}
\end{equation}
After extracting the electron part from Eq.
\eqref{spectaltensor} and applying Eq. \eqref{spectralfunctiona}, the
spectral function can be written as:
\begin{equation}
\rho_{s s'} =
 -\frac{1}{\pi}\sum_{\alpha} (G_{s \alpha}^{+} \Sigma{''}_{\alpha} G_{\alpha
s'}^{-} + F_{s \alpha}^{+} \Sigma{''}_{\alpha} F_{\alpha
s'}^{-}),
\label{spectralfunction}
\end{equation}
Using Eq. \eqref{spectralfunction}, the tunneling current of Eq.
\eqref{conductance} can be recast into the form
\begin{equation}
\sigma = \sum_{t \alpha} T_{t \alpha}, \label{transition}
\end{equation}
where
\begin{widetext}
\begin{equation}
 T_{t \alpha}  = -\frac{2 e^2 }{ \hbar}\sum_{t' s s'}
 \rho_{tt'}(E_F)
 V_{t's}(G^{+}_{s\alpha}\Sigma{''}_{\alpha}G^{-}_{\alpha
   s'} + F^{+}_{s\alpha}\Sigma{''}_{\alpha}F^{-}_{\alpha
   s'})V_{s't}^{\dagger},
\label{partial}
\end{equation}
\end{widetext}
and the Green's function and the self-energy are evaluated at energy $E
= E_F + e V_b.$ Eqs. \eqref{transition} and \eqref{partial} are an
extension of the Landauer-B\"uttiker formula for tunneling across
nanostructures (see, e.g., Ref.~\onlinecite{Meir}), and represent a
reformulation of Refs.~ \onlinecite{Fisher} and
\onlinecite{Frederiksen}.  By comparing Eqs. \eqref{spectralfunction}
and \eqref{partial}, we see that if the tip makes contact with only a
single surface atom orbital, e.g., a Bi-$p_z$ orbital, then the
tunneling current is directly proportional to the LDOS {\it of that
orbital}. In particular, the tunneling current bears in general 
no such simple relationship to the quantity of most interest, namely, the 
LDOS on the CuO$_2$ plane. Obviously,
the tunneling formalism of Eq.  \eqref{partial} must be further
elaborated in order to find the relation between the
interesting LDOSs and the tunneling spectrum.

\subsubsection{Tunneling channels, filter function and tunneling
  matrix element}

The experimental STM spectra in the cuprates have to date been mostly 
compared to the electronic LDOS of the superconducting cuprate layer, 
especially the LDOS of the Cu-$d_{x^2-y^2}$ orbital. The discrepancies 
between the spectra and the LDOS are then ascribed to `tunneling matrix 
elements' or `filtering functions' \cite{Balatsky}. The former refers to 
the general problem of modeling spectroscopies, where the signal is 
distorted by the spectroscopic process, and may even vanish due to the 
presence of selection rules. The latter term refers to how the states of 
electrons (or quasiparticles) from the initial state within the 
superconducting layers are modified when traveling through the oxide 
overlayers before reaching the tip. Eq. \eqref{partial} above accounts 
fully for the tunneling process, and it can be reformulated to reveal, for 
example, the filtering effect more clearly. For this purpose, it is 
convenient to the denote various orbitals as follows: $s$ and $s'$ for the 
orbitals of the sample surface, which overlap with the tip orbital $t$; 
$f$ and $f'$ for the orbitals of the filter layers, BiO and SrO; $c$ and 
$c'$ for orbitals in the cuprate layer; and, $\alpha$ for any orbital that 
is singled out, which in our case usually will be an orbital in the 
cuprate layer. Denoting the Green's function for the filter layers
decoupled from the rest of the system by $G^{0+}_{sf}$, 
and the matrix elements within the cuprate layer
in the coupled system by $G^{+}_{c\alpha}$, application of Dyson's
equation to $G^{+}_{s\alpha}$ yields
\begin{displaymath}
G^{+}_{s\alpha}  =  G^{0+}_{sf}V_{fc}G^{+}_{c\alpha}~~
\textrm{and}~~
F^{+}_{s\alpha}  =  G^{0+}_{sf}V_{fc}F^{+}_{c\alpha}
\end{displaymath}
Hence, Eq. \eqref{partial} can be written as
\begin{widetext}
\begin{equation}
T_{t \alpha} =
-\frac{2 e^2 }{ \hbar}\sum_{t' c c'}
 \rho_{tt'}(E_F)M_{t'c}(G^{+}_{c\alpha}\Sigma{''}_{\alpha}G^{-}_{\alpha
   c'}+F^{+}_{c\alpha}\Sigma{''}_{\alpha}F^{-}_{\alpha
   c'})M_{c't}^{\dagger}
\label{cupraspectral}
\end{equation}
\end{widetext}
where
\begin{equation}
M_{tc} = V_{ts}G^{0+}_{sf}V_{fc},
\label{filter}
\end{equation}
which gives the filtering amplitude between the cuprate layer and the
tip, and constitutes a multiband generalization of filtering function
of Ref. \onlinecite{Balatsky}.  Similarly, the matrix element of the
density of states operator $\rho_{cc'}$ within the cuprate plane can
be recovered in terms of the spectral function:
\begin{equation}
\sigma =
 \frac{2 \pi e^2 }{\hbar}\sum_{tt' c c'}
 \rho_{tt'}(E_F)M_{t'c}\rho_{cc'}(E_F+eV)M_{c't}^{\dagger},
\label{cupraldos}
\end{equation}

Eqs.~\eqref{cupraspectral}-\eqref{cupraldos} show a number of
interesting aspects of the tunneling process as follows.\\ 
\indent (1) Since applying the {\it filtering matrix element} $M_{tc}$, which
describes the effect of the BiO and SrO overlayers, involves
$M$ and $M^{\dagger}$, interference effects will occur between various
paths to the tip from the cuprate layers through the filter layer;\\
\indent (2) The partial current terms in \eqref{cupraldos} under the
summation are proportional to elements of the density matrix confined
to the cuprate layer. Only orbitals with a notable overlap with the
$p_z$ orbital of the apical oxygen on the SrO layer will give a
significant contribution to the total current;\\ 
\indent (3) The partial elements of the {\it spectral function}
\begin{equation}
\rho_{c c'  \alpha}= -\frac{1}{\pi}
(G^{+}_{c\alpha}\Sigma{''}_{\alpha}G^{-}_{\alpha
   c'}+F^{+}_{c\alpha}\Sigma{''}_{\alpha}F^{-}_{\alpha c'})
\label{singlespec}
\end{equation}
extracted from Eq. \eqref{cupraspectral} show which orbitals $\alpha$
 contribute to the chosen element of the density matrix $\rho_{cc'}.$
 Furthermore, the current contribution $T_{t \alpha}$ between the tip
 can be divided into regular and anomalous terms $T^{R}_{t \alpha}$
 and $T^{A}_{t \alpha},$ respectively \cite{footdecomposition}.

Since the filter layers are insulating at low energies, these layers will 
give little structure to the spectrum at low bias voltages, so that the 
structure of the spectrum is mainly controlled by the matrix elements 
$\rho_{cc'}$, and in this sense the spectrum is a 
filtered mapping of the LDOS of the cuprate orbitals. We will show however 
that the Cu-$d_{x^2-y^2}$ orbitals right below the tip do not enter the 
spectrum through Eq.~\eqref{cupraldos} since their overlap with the 
relevant orbitals of the SrO layer is zero. Instead, Cu-$d_{z^2}$ has a 
large overlap with $p_{z}$ of the apical oxygen and hence these orbitals 
of the Cu atoms play a dominant role in the tunneling spectrum.

The detailed contribution of any specific orbital $\alpha$ can be 
extracted from Eq.  \eqref{cupraspectral}. The regular and anomalous 
matrix elements of the spectral function, 
$G^{+}_{c\alpha}\Sigma{''}_{\alpha}G^{-}_{\alpha c'}$ and 
$F^{+}_{c\alpha}\Sigma{''}_{\alpha}F^{-}_{\alpha c'}$, describe 
propagation 
of electrons or holes within the cuprate layer from orbital $\alpha$ to 
the orbitals $c$ and $c'$. The latter orbitals act as ``gates'' between 
the cuprate layer and the filter layer.  For example, if $\alpha$ is 
$d_{x^2-y^2}$ of a Cu atom and $c$ and $c'$ are $d_{z^2}$ orbitals, which 
strongly overlap with the filter layer, the matrix element filtered by $M$ 
and $M^{\dagger}$ gives the contribution of a specific $d_{x^2-y^2}$ 
orbital to the total tunneling spectrum. 
Note that in the superconducting state the anomalous
matrix elements of the spectral function must also be considered.  
$F_{\alpha
  \beta}(\tau)$ involves the creation of an electron with spin
up coupled to the annihilation of a hole with spin down given by $\langle
c^{\dagger}_{\beta \downarrow}(\tau) c^{\dagger}_{\alpha \uparrow}(0)
\rangle,$ and thus describes the formation
and breakup of Cooper pairs as shown in Appendix B. The decomposition of 
Eqs.~\eqref{cupraspectral}-\eqref{cupraldos}
are, in fact, a generalization of the tunneling channel
approach to transport through one-molecule electronic components
\cite{Magoga} and STM of adsorbate molecules \cite{Sautet, Niemi}.  In
the present context, the ``tunneling path'' analysis gives us the
``origin'' of the signal, since $G^{+}_{c\alpha}\Sigma{''}_{\alpha}
G^{-}_{\alpha c'}$ gives the probability of propagation between
orbitals $\alpha$ and $c.$

\section{Results}

\subsection{Topographic maps}

\begin{figure}[h]
\centering
    \includegraphics[width=0.50\textwidth]{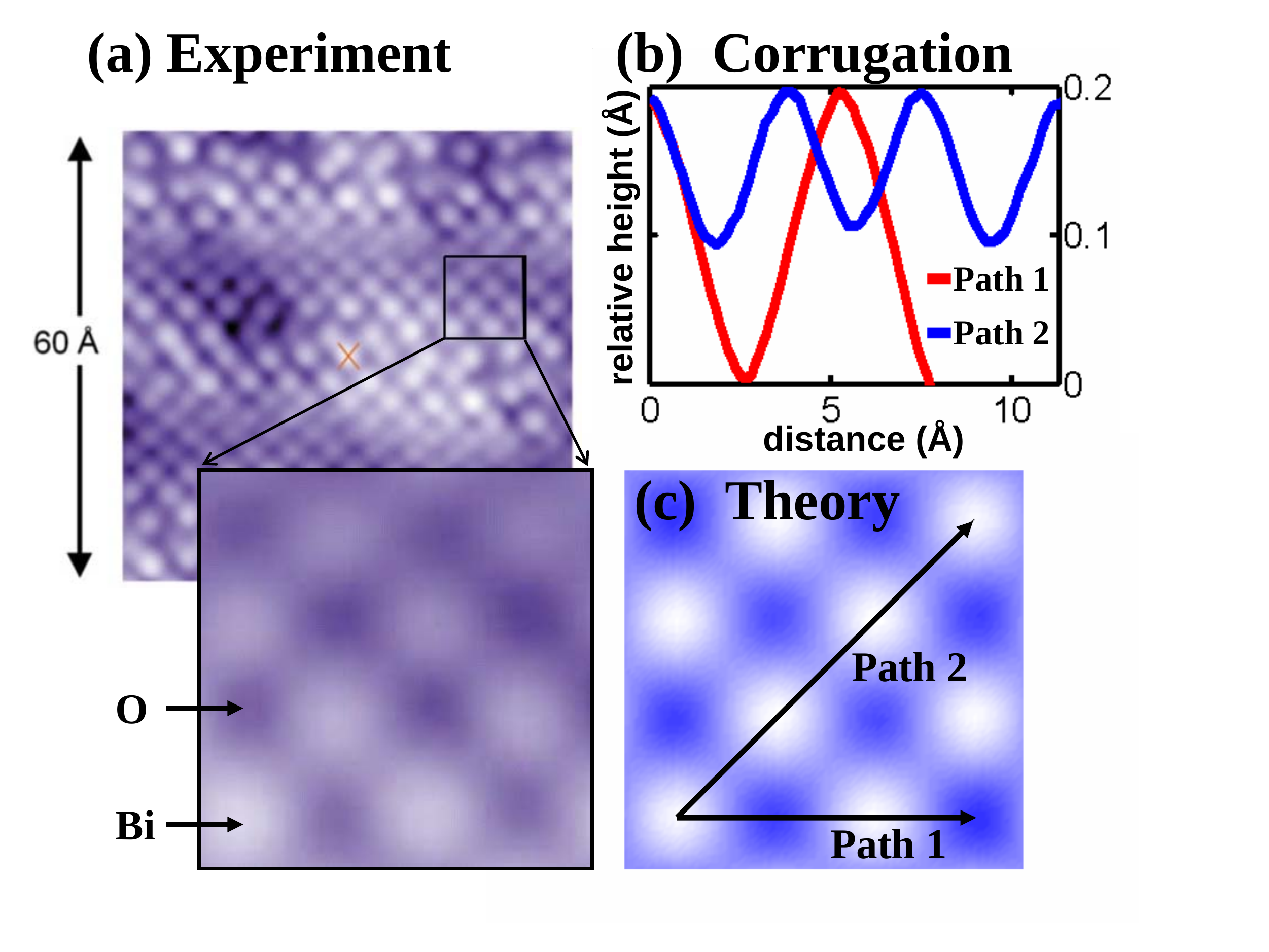}
  \caption{ (a) Typical experimental topographical STM map after
Ref. \onlinecite{Pan}. (b) The computed corrugation of two STM line
scans and (c) theoretically predicted topographic map. The two paths
are shown in (c) by arrows.  }
  \label{exptheo}
\end{figure}

We discuss first the topographic STM map, i.e., the constant current
surface for a tip scanning across the sample surface. The computed
topographic map is very robust against changes in measuring parameters
such as the bias voltage or the tip-surface distance. Figure
\ref{exptheo} compares the calculated and typical experimental
results. Furthermore, corrugation along two paths of line scan is
shown in Fig. \ref{exptheo} (b). The Bi atoms are seen as bright
spots, while the surface oxygens are dark due to very low current
coming through these surface atoms. We will see in connection with the
analysis of the tunneling channels below that the apical oxygens act
as the primary gate for passing electrons from the CuO$_2$ layers up
to the surface BiO layer. Accordingly, the Bi atoms appear bright
because there exists an easy channel between the surface Bi atoms and
the apical oxygens below via the Bi~$p_z$ orbitals.  On the other
hand, the oxygens in the surface layer are dark because the $p_{x,y}$
orbitals of O(Bi) are orthogonal to the (assumed) $s$-symmetry of the
tip, while the O(Bi) $p_{z}$ orbitals are relatively weakly coupled to
the $p_{z}$ of the apical oxygen as discussed below in connection with
Fig. \ref{channels08}.

\subsection{Tunneling spectra}

\begin{figure}[h]
\centering
    \includegraphics[width=0.50\textwidth]{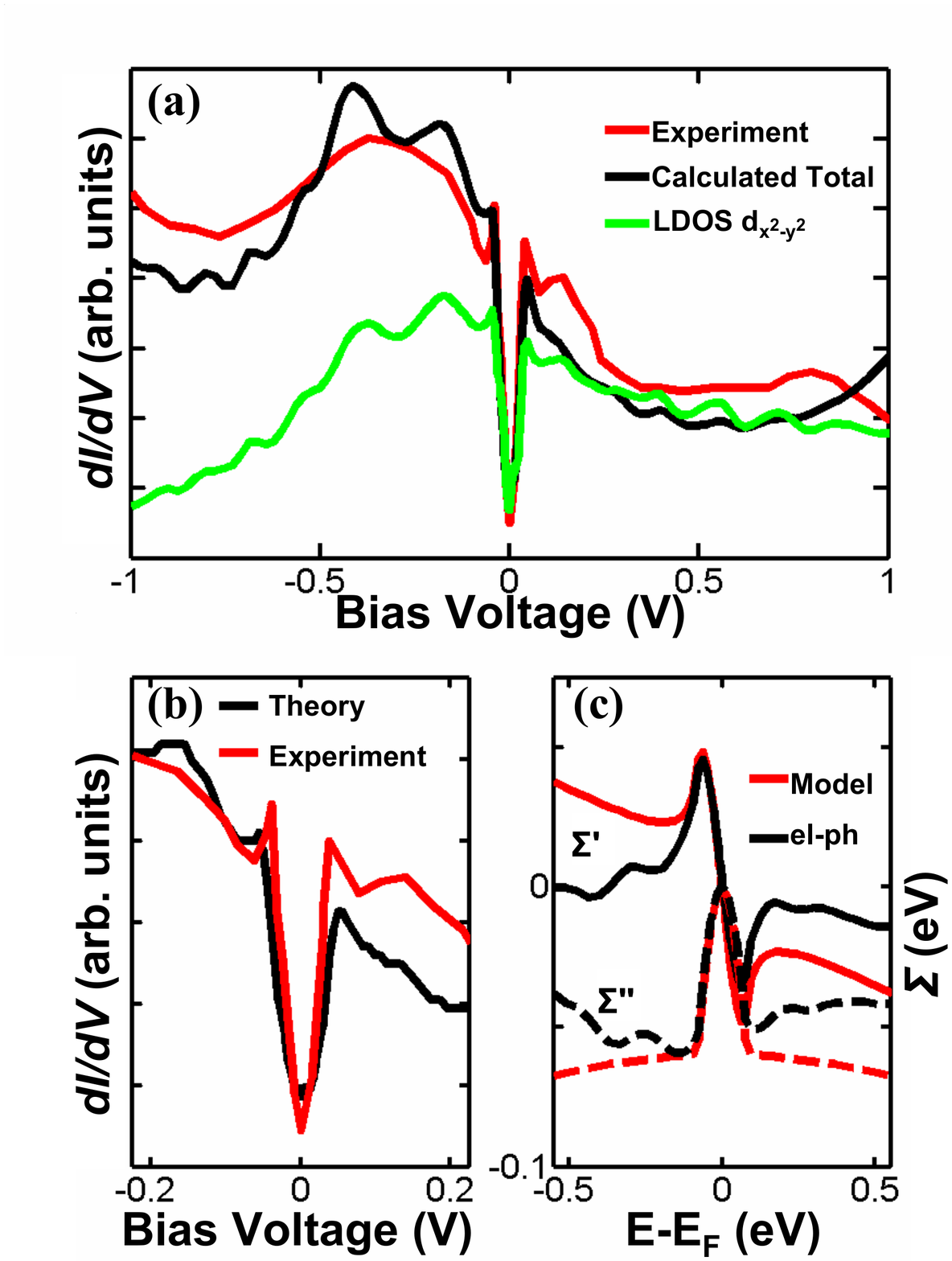}
  \caption{(color online) (a) A typical experimental tunneling
    spectrum (red line) from Bi2212 (after Ref. \onlinecite{McElroy})
    is compared with the calculated spectrum (black).
     The green curve shows the LDOS
    of the Cu-$d_{x^2-y^2}$. (b) Expanded view of the experimental and
    calculated spectrum in the low energy region. (c) Comparison of
    the model self-energy (Eq. \eqref{sigmaeinstein}) assumed for the
    Cu-$d_{x^2-y^2}$ orbitals and the self-energy from the convolution
    of a Debye-type phonon spectrum and the LDOS of Cu-$d_{x^2-y^2}$
    (Eq. \eqref{self}) as discussed in Appendix A.  }
  \label{pristine}
\end{figure}

Fig. \ref{pristine} (a) compares a typical experimental (red line) STS
spectrum\cite{McElroy} to the calculated one (black line). The overall
agreement between theory and experiment is seen to be good, although
the VHSs are seen as separate structures in the calculated curve
\cite{footnote2,MSB}. 
The agreement also extends to the low energy region
shown in Fig. \ref{pristine}(b), where the width and positions of the
coherence peaks is reproduced reasonably well.\cite{footkpoints} The
tendancy for increasing intensity towards negative bias is seen in
both measurements and computations. This is in sharp contrast to the
shape of the LDOS of Cu-$d_{x^2-y^2}$ orbital (green curve). As
emphasized in Ref. \cite{NLMB}, this remarkable asymmetry of the
spectrum between positive and negative bias voltages reflects the
opening up of channels other than Cu-$d_{x^2-y^2}$, especially of
Cu-$d_{z^2}$, as one goes to high negative bias. This asymmetry thus
appears naturally within our conventional picture and cannot be taken
to be a hallmark of strong correlation effects as has been thought to
be the case.

There has been considerable interest in understanding the coupling of
electrons to bosonic modes in the so-called `low-energy kink' region
within $\sim\pm 100$~meV of the Fermi level. In particular, the
peak-dip-hump structure seen in the experimental spectrum in
Fig. \ref{pristine}(b) is generally believed to be the result of the
coupling of electronic degrees of freedom to a collective mode
(Refs. \onlinecite{kinks,hoogenboom,VHS}).  Fig. \ref{pristine}(b)
shows that the peak-dip-hump feature can be described by our simple
self-energy correction discussed in Appendix A. This point however
requires further study, including an analysis of how this feature
evolves with doping.

\subsection{Selection rules}

\begin{figure}[h]
\centering
    \includegraphics[width=0.50\textwidth]{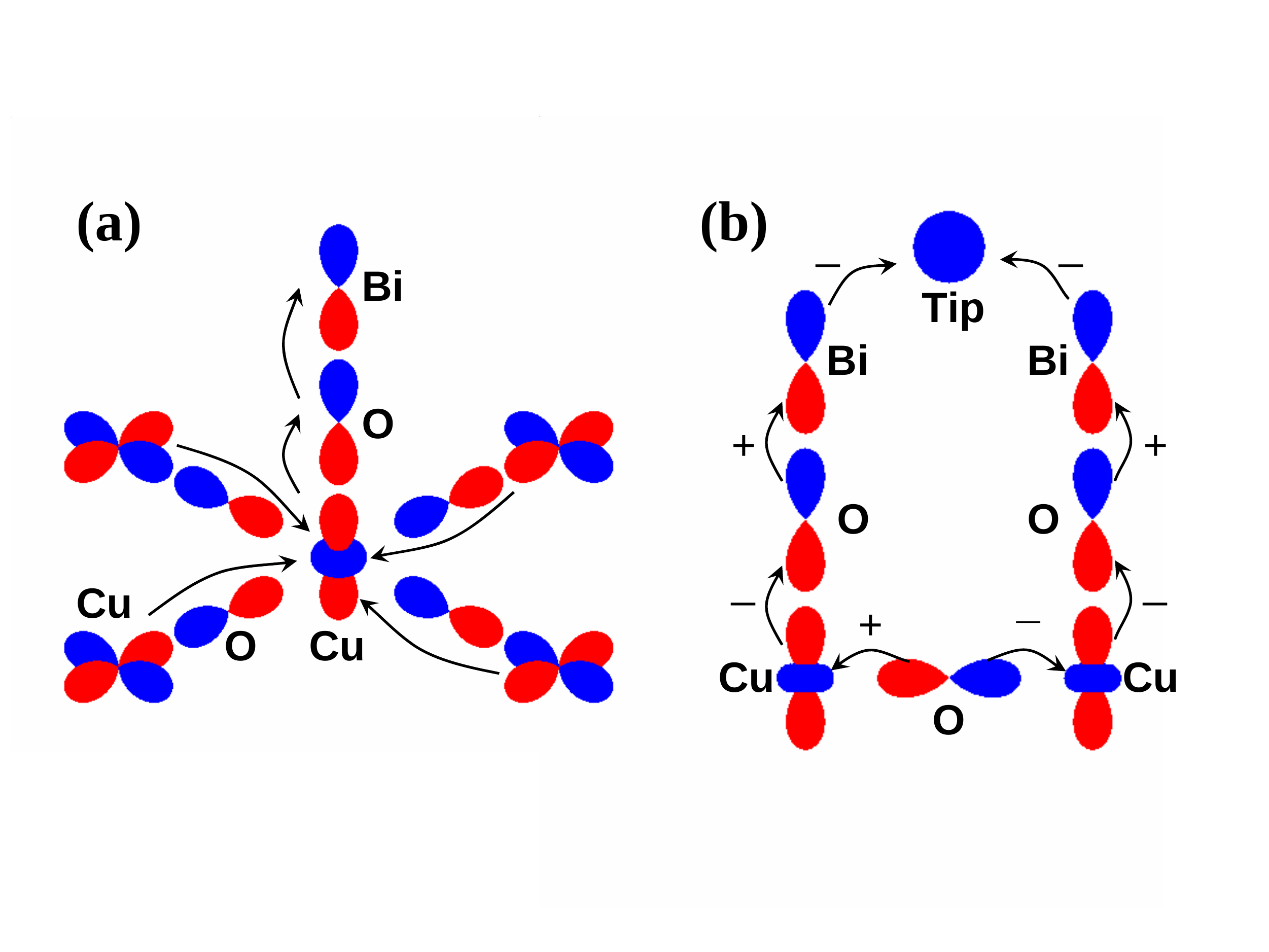}
  \caption{(color online) 
(a) Dominant tunneling channel from the
    cuprate layer, from Cu $d_{x^2-y^2}$ orbitals through the neighboring
    Cu $d_{z^2}$ to Bi $p_{x}$ to the tip. (b) An oxygen atom
    in the cuprate layer is invisible to a STM tip right above, since
    the paths through Cu$_{1}$ and Cu$_{2}$ interfere
    destructively.
}
  \label{channels08}
\end{figure}

The filter function $M_{tc}$ controls {\it selection rules} dictated
by matching of the symmetry properties of the cuprate layer, filter
layers and the tip.  A closer examination of $M_{tc}$ reveals that
strong tunneling through the apical oxygen layer is associated with a
matching of the symmetry of the cuprate layer wave function to that of
the apical O-$p_z$. The key is the relative symmetry of the wave
functions with respect to the axis of tunneling: An `odd' wave
function, e.g., the Cu-$d_{x^2-y^2}$ has zero overlap with an `even'
wave function such as O-$p_z$. In contrast, two orbitals with the same
symmetry couple more strongly. Accordingly, the $p_z$ of the apical
oxygen and the Cu-$d_{z^2}$ possess large overlap, while
Cu-$d_{x^2-y^2}$ has zero overlap with any $s$- or $p$-orbital of the
apical oxygen. This is the reason that direct tunneling is forbidden
between Cu-$d_{x^2-y^2}$ and the $s$-wave symmetric tip through the
filter layer. Hence, $M_{tc}$ functions here are consistent with the
filter function of Ref. \onlinecite{Balatsky}. Similarly, coupling
between an $s$-wave tip and the $p_x$ and $p_y$ orbitals of the Bi
atom lying directly below the tip is forbidden.  Therefore, within the
filter layer, the main `vertical' overlap is between the $p_z$
orbitals of Bi and apical oxygen, and these orbitals indeed are found
to provide the main channel through the filter layers as depicted in
Fig. \ref{channels08}(a). We find additional relatively small
contributions from the on-site Bi-$s$-orbital and $p$-orbitals of the
surrounding Bi and O(Bi) atoms, but such 'background' contributions to
the current do not seem to be dominated by any particular channel.

Figure~\ref{channels08}(b) illustrates another example of a
  symmetry-forbidden tunneling path, where the tip is centered between
  two surface Bi's, i.e. on the top of an oxygen of the
  cuprate layer. Since we assume an $s$-wave
  tip with negative hopping integrals to the nearby Bi atoms, when 
we follow either path up to the Cu-$d_{z^2}$ orbitals,
  the signs of the hopping integrals are identical. However, the
  O-$p_{x}$ orbital between the two Cu atoms changes sign from one Cu
  to the other. This gives the two paths from O-$p_x$ to the
  $s$-wave tip an opposite phase leading to destructive
  interference between the paths, making the O atom invisible.
However, if the $s$-wave tip is replaced by one with, e.g, 
$p_x$ symmetry, the oxygen would become visible and a weaker
signal would appear from the neighboring $d_{x^2-y^2}$ orbitals.
Experimentally, this could be accomplished by functionalizing
the tip by attaching a suitable molecule to the tip. A similar procedure 
has been used to obtain a contrast inversion for CO molecules adsorbed 
on
a Cu surface \cite{Rieder, Niemi1}.

\subsection{Tunneling channels}

\begin{figure}[h]
\centering
    \includegraphics[width=0.50\textwidth]{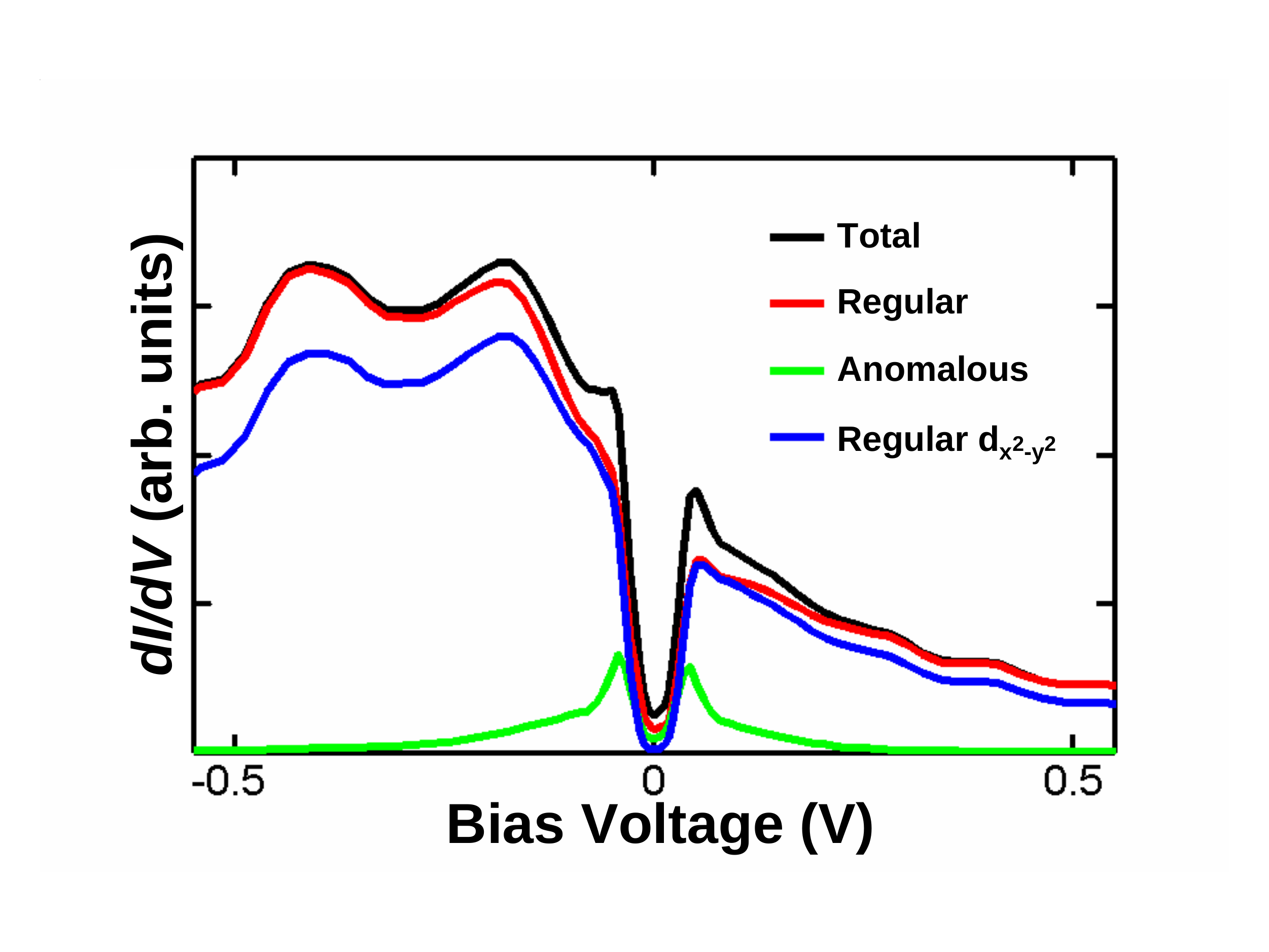}
  \caption{(color online) Partial spectrum with $c=c'=d_{z^2}$ in $M_{tc}.$
The regular (red line) and anomalous (green line) components are
shown together with the total contribution of the two parts (solid black). 
Blue curve shows the corresponding regular Cu-$d_{x^2-y^2}$ contributions.
}
  \label{partspectra1}
\end{figure}

\begin{figure}[h]
\centering
    \includegraphics[width=0.50\textwidth]{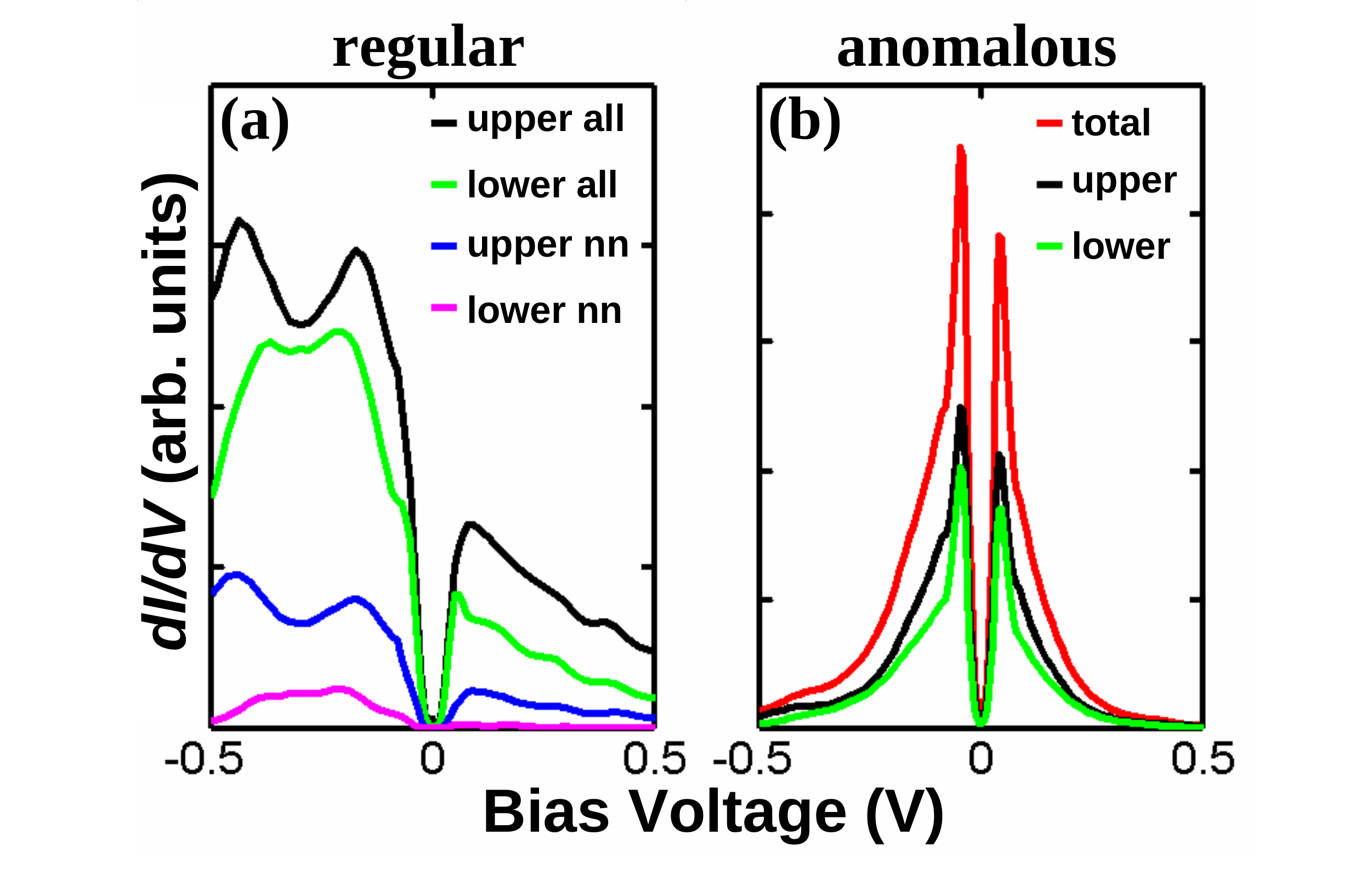}
  \caption{(color online) (a) Various contributions to tunneling
spectrum from the regular matrix elements (assuming $c=c'=d_{z^2}$ in
$M_{tc}$), $T^{R}_{t \alpha}$, of Cu-$d_{x^2-y^2}$ orbitals of upper
and lower CuO$_2$ layer. Contributions from the nearest neighbor (nn)
Cu atoms in the upper and lower layer are shown.  (b) Same as (a),
except this panel refers to the contributions from the anomalous
matrix elements, $T^{A}_{t \alpha}$.  }
  \label{partspectra2}
\end{figure}

The origin of the current from the cuprate layer can be understood by
inspecting the individual terms of Eq.  \eqref{singlespec}, which we
refer to as 'tunneling channels', i.e., from the regular and anomalous
elements $G_{c \alpha}^{+} \Sigma{''}_{\alpha} G_{\alpha c'}^{-}$ and
$F_{c \alpha}^{+} \Sigma{''}_{\alpha} F_{\alpha c'}^{-}$, of the
Green's function. [Although tunneling channels are a normal state
property, the anomalous matrix elements play an important role in
generating the coherence peaks and thus are relevant more generally.]
For simplicity, we assume that the tip is right above a Bi atom. The
dominant element of the filter function $M_{t,c}$ is then between the
tip orbital and the $d_{z^2}$ orbital of the upper layer Cu atom lying
beneath the surface Bi atom, so we take $c=c'=$ Cu-$d_{z^2}$ in
results shown in Figs. \ref{partspectra1} and
\ref{partspectra2}. Fig. \ref{partspectra1} shows the relative
contributions of the regular and anomalous matrix elements. The near
Fermi energy current is primarily associated with the $d_{x^2-y^2}$
matrix elements. While the regular matrix elements of Cu-$d_{x^2-y^2}$
are almost solely responsible for the spectrum at energies around the
VHSs, the {\it anomalous} elements determine the features around the
gap region, especially the {\it coherence
peaks}. Fig. \ref{partspectra1} shows that coherence peaks are
inherited from the anomalous and not the regular part of the Green's
function, and reflect physically the effects of non-conservation of
the number of electrons near the gap region.

In Fig. \ref{partspectra2}, the current of $d_{x^2-y^2}$ character is 
further broken down into contributions from various neighbors of the 
central Cu atom of the first and second CuO$_2$ layer away from the free 
surface. We see in panel (a) that the upper CuO$_2$ layer is more 
important than the lower one, but that the upper layer is by no means 
dominant. It seems that the coupling between the tip and the lower layer 
is strengthened via the relatively large overlap between the $d_{z^2}$ 
orbitals of the central Cu atoms of the two layers, which opens an 
important interlayer channel. The $d_{x^2-y^2}$ orbitals of the two layers 
mix not only to induce the well-known bilayer splitting in Bi2212, but 
also play a significant role in the flow of current to the tip from the 
lower cuprate layer.

It can be seen from Fig. \ref{partspectra2} (a) that the $d_{x^2-y^2}$
orbitals of the four nearest-neighbor Cu atoms of the central Cu give
a significant contribution to the total spectrum, but that this
amounts to only about one third of the contribution from all
$d_{x^2-y^2}$ terms from the upper layer. Due to the non-local nature
of the Bloch-states within the cuprate layers, it is clear then that
the total signal involves long range contributions, and attributing
the spectrum merely to the four nearest neighbor Cu atoms provides
only a rough approximation.

Anomalous contributions are considered in Fig. \ref{partspectra2} (b). 
Here, the upper and lower layers give an almost equally large 
contribution, indicating that coherence peaks also are not all that local 
in character. Notably, we find a finite onsite {\it anomalous} 
contribution of $d_{x^2-y^2}$ even though the regular term is zero.  
This can be understood with reference to 
Eq. \eqref{fdelta}. Consider the term $$F_{z^2 d}= G_{e,z^2 
(d+x_i)}^{0}\Delta_{(d+x_i) d} G_{h,d d}, $$ where $d$ is shorthand for 
$d_{x^2-y^2}$ of the central Cu and $d+x_i$ is $d_{x^2-y^2}$ of the 
neighboring Cu in either $x$- or $y$-direction.  Clearly, $G_{e,z^2 
(d+x_i)}^{0}$ transforms under rotations of $\frac{\pi}{2}$ in the same 
way 
as $\Delta_{(d+x_i) d},$ and since $G_{h,d d}$ is an onsite term, the 
combination is invariant. Hence the four terms in the sum over the 
neighbors are equal, yielding a non-zero onsite term.

We emphasize that the {\it anomalous contribution} of the four neighboring 
Cu atoms is quite small. Let us consider the term $$F_{z^2 (d+x_{i})}= 
G_{e,z^2d}^{0}\Delta_{d (d+x_i)} G_{h,(d+x_i) (d+x_i)}.$$ Due to 
symmetry, $G_{e,z^2d}^{0} = 0,$ and thus this term vanishes. However, there 
are terms like $$F_{z^2 (d+x_{i})}=G_{e,z^2 (d+2x_i)}^{0}\Delta_{(d+2x_i) 
(d+x_i)} G_{h,(d+x_i)  (d+x_i)}$$ which do not vanish, but are very 
small, since $G_{e,z^2 (d+2x_i)}^{0}$ is a relatively small term. A similar 
analysis can be carried out for the second and third neighbors.  The 
second nearest neighbors, which lie along the nodal direction in k-space, 
give the largest single contribution, although this contribution is not 
dominant. The third neighbor contribution is a little larger than the 
onsite contribution.

\section{Further Comments}

\subsection{Symmetry Analysis}

\begin{figure}[h]
\centering
    \includegraphics[width=0.45\textwidth]{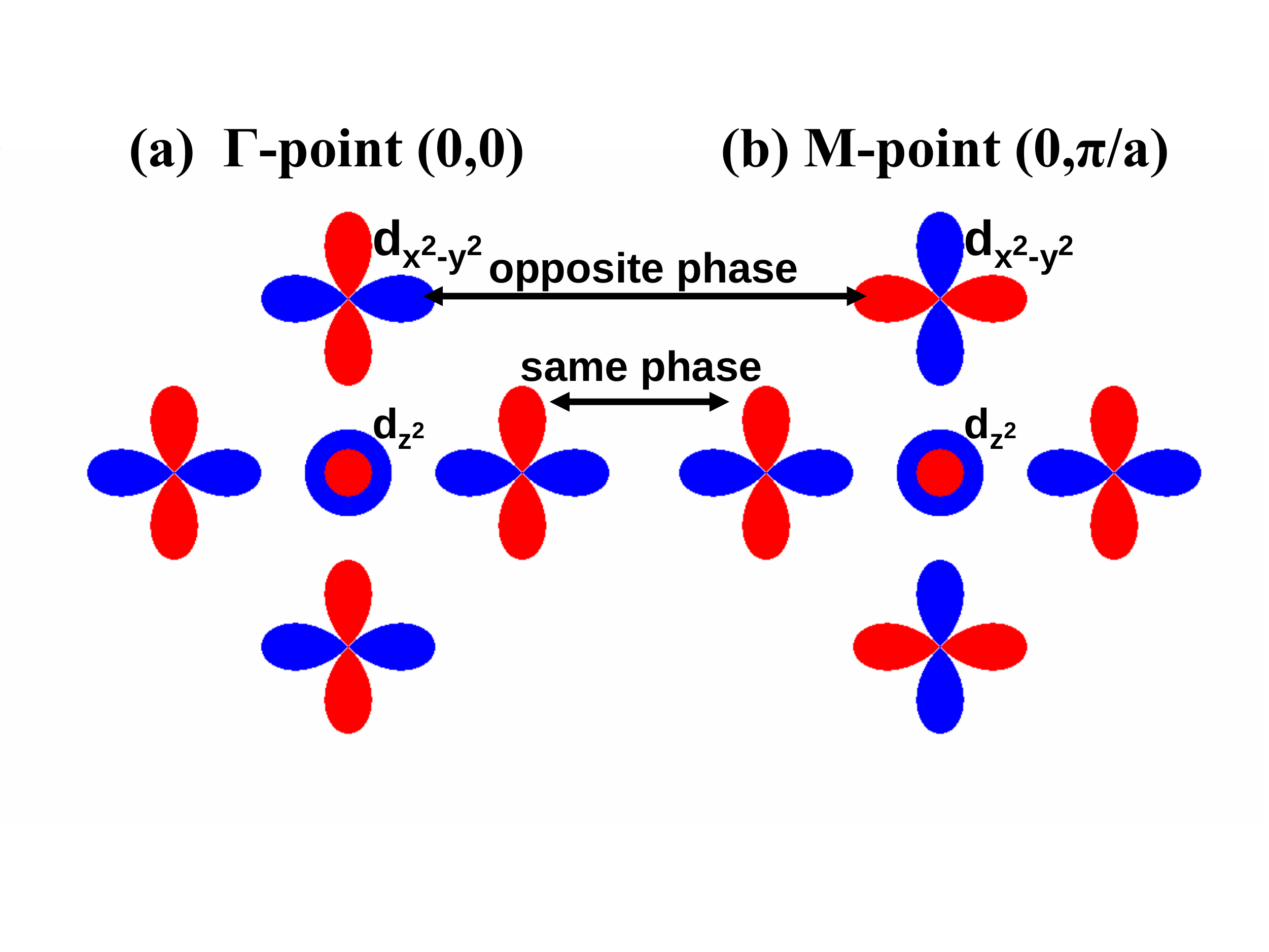}
  \caption{(color online) Relative phases of the central $d_{z^2}$
    orbital and the neighboring $d_{x^{2}-y^{2}}$ orbitals at the
    $\Gamma$ point (a) and at the $M$ point (b).}  
  \label{phases}
\end{figure}

The selection rules can be formalized using group theoretical arguments 
related to the filtering function. \cite{Balatsky}.  For example, in order 
to explain the dominance of the $d_{x^2-y^2}$ orbitals of the four 
neighboring Cu atoms, considering representations of the two-dimensional 
$C_{4v}$ group, the d-orbitals $\vert d_{x^2-y^2},i\rangle$ of the site 
$i$ 
participate in eigenfunctions of the system as a linear combination 
\begin{displaymath} \displaystyle{\sum_{i}} e^{-i\mathbf{k}\cdot 
\mathbf{R}_i} \vert d_{x^2-y^2},i\rangle. \end{displaymath} This 
combination of the four neighboring orbitals at $(0,\pm \pi)$ and $(\pm 
\pi, 0)$ belongs to the same representation of $C_{4v}$ as the $4s$ and 
$d_{z^2}$ orbitals of the central Cu atom (see Fig. \ref{phases}), as well 
as the $p_z$ orbitals of the apical oxygen and the surface Bi atom. At 
this k-point, the phase difference between the lattice sites causes all 
the d-orbital lobes pointing towards the central atom to have the same 
sign.  Hence, this combination yields a large off-diagonal element overlap 
with the surface $p_z$-orbital, and a dominant tunneling contribution 
around the gap. Similar arguments can be applied to understand 
contributions from other farther out atoms. An example was given in 
Fig.~\ref{channels08}(b) above where the position of the tip and the 
symmetry of the relevant orbital strongly influence the visibility of an 
atom.

\subsection{Electron extraction/injection}

To relate the tunneling current to the LDOS of the cuprate layer, we
have introduced the concept of tunneling paths through
Eq. \eqref{cupraspectral}, which implies that each path {\it begins or
ends on a particular atomic orbital}. This non-intuitive concept
requires some comment. In reality, the current flows through the
sample with each electron ejected to the tip being replaced by an
electron from a distant counterelectrode. For a simple system, such as
a nanostructure, non-equilibrium Green's function formalism with two
`leads' closing a current circuit have been invoked (see, e.g., Ref.
\onlinecite{Meir}). Tersoff-Hamann (TH) or Todorov-Pendry (TP)
approach, on the other hand, assumes that the current is composed of a
series of tunneling events\cite{THTPfoot}, and that the replacement of
electrons at the counterelectrode has a negligible effect on the
tunneling process.  Since the current in STS is of the order of
$10-100pA$, there is only about one electron each $1-10ns$ which flows
across the sample, justifying the assumptions underlying TH/TP
approach. Both TH and TP are based on calculating individual tunneling
events in a LEED-like formalism\cite{ShenRMP}. Due to the finite
$\Sigma"$, an electron created on a particular atom will have only a
finite probability of escaping to the tunneling tip, and
Eq. \eqref{cupraspectral} shows how to add up the contribution of all
these tunneling processes in terms of the equilibrium LDOS of the
sample.

\section{Conclusions}

We have presented a comprehensive framework for modeling the STS spectra 
from the normal as well as the superconducting state of complex materials 
in a material-specific manner. Our formulation makes transparent the 
connection between the LDOS and the STS spectrum or the nature of the 
tunneling 'matrix element', and it is cast in a form that reveals the 
filtering effect of the overlayers separating the tip and the layers of 
interest. Our decomposition of the tunneling current into contributions 
from individual local orbitals allows us to identify important 'tunneling 
channels' or paths through which current reaches the STM tip in the 
system. Our analysis highlights the importance of anomalous terms of the 
Green's function, which account for the formation and breaking up of 
Cooper pairs, and how such terms affect the STS spectrum.

We apply the formalism to the specific case of Bi2212. Mismatch of 
symmetry between orbitals on adjacent atoms, or between the tip and the 
sample orbitals, is shown to severely restrict the corresponding 
contribution to the tunneling current. For these reasons, the contribution 
from Cu-$d_{x^2-y^2}$ orbitals comes not directly from the Cu-atom lying 
right below the Bi atom, but from a fourfold symmetric indirect route 
involving the four nearest-neighbors of the central Cu as well as 
longer range background from farther out Cu-$d_{x^2-y^2}$ 
orbitals. In the superconducting state, the coherence peaks of the 
spectrum are shown to be dominated by the anomalous spectral terms, which 
also are found not to be all that localized around the central Cu atom. In 
particular, we find a small anomalous on-site term and a practically 
vanishing first nearest neighbor contribution, with most of the anomalous 
contribution arising from the second neighbors and beyond.

We have concentrated in this study on the large hole doping regime of the 
cuprates where a homogeneous electronic Fermi liquid phase is consistent 
with most experiments.  The fact that we have obtained good overall 
agreement between our computations and the measurements, especially with 
respect to the pronounced asymmetry of the spectrum between positive and 
negative bias voltages, indicates that this remarkable asymmetry can be 
understood more or less within our conventional picture without the need 
for invoking exotic mechanisms. At lower dopings, strong correlation 
effects including the possible presence of competing orders or 
inhomogeneous electronic states (nanoscale phase separation) would need to 
be taken into account. However, the present framework can be extended 
fairly straightforwardly through the addition of Hubbard terms in the 
Hamiltonian to provide a viable scheme for investigating the tunneling 
response throughout the phase diagram of the cuprates and other complex 
materials, including the modeling of effects of impurities and dopant 
atoms in the system.

{\bf Acknowledgments}

This work is supported by the US Department of Energy, Office of
Science, Basic Energy Sciences contract DE-FG02-07ER46352, and
benefited from the allocation of supercomputer time at NERSC,
Northeastern University's Advanced Scientific Computation Center
(ASCC), and the Institute of Advanced Computing, Tampere.  RSM's work
has been partially funded by the Marie Curie Grant PIIF-GA-2008-220790
SOQCS. I.S. would like to thank the Wihuri Foundation for financial
support. Conversations with Jose Lorenzana and Matti Lindroos are
gratefully acknowledged.

\appendix
\section{Boson-electron coupling}

In the vicinity of the Fermi energy, dispersion anomalies are found in 
ARPES spectra arising from coupling of electronic degrees of freedom to 
phonons and/or magnetic modes, often giving the appearance of a 
peak-dip-hump feature\cite{kinks}.  These boson-electron couplings also 
strongly affect the STS spectrum\cite{VHS}.  This appendix discusses a 
model self-energy for describing such anomalies.

A significant contribution to the electron-phonon coupling is
associated with modulation of the electronic hopping integrals by the
phonons.  The generalized coordinate of atomic displacement in $q$-basis
is quantized in the standard way:
\begin{displaymath}
Q_{q}= \sqrt{\frac{\hbar}{2 \Omega_q}}\left(a_q + a^{\dagger}_q  \right),
\end{displaymath}
where $a_{q} (a^{\dagger}_{q})$ is the annihilation (creation) operator of the
phonon mode $q$, and $\Omega_q$ is the frequency of the mode. However,
the most natural way to couple this to real-space tight-binding basis is
to make a transformation to
the basis of real space displacement of atom $\mu$ in the
following way:
\begin{displaymath} 
\hat{u}_{\mu} = \langle \mu \vert q \rangle Q_{q},
\end{displaymath}
where Einstein summation over phonon modes $q$ is implicit.  Note, that
$\mu$ is a composite index denoting both the index of an atom and the
direction of displacement.

Consequently, in tight-binding basis, this gives rise to a term in the 
Hamiltonian of the form
\begin{displaymath}
  H^{el-vib} = \frac{1}{\sqrt{m_{\mu}}}
\frac{\partial V_{\alpha \delta}}{\partial R_{\mu}} \hat{u}_{\mu}
 c^{\dagger}_{\alpha} c_{\beta}=
\Gamma_{\mu}^{\alpha \delta} \hat{u}_{\mu} c^{\dagger}_{\alpha} c_{\beta}
\end{displaymath}
where $V_{\alpha \delta}$ is the hopping integral between orbitals
$\alpha$ and $\delta$, $ R_{\mu}$ is the coordinate of atom $\mu$.

This coupling can be embedded into the electronic
Hamiltonian as an energy dependent self-energy.  Following the
arguments of Ref.  \onlinecite{paulsson-05}, the general form of
self-energy is written as:
\begin{eqnarray}
\Sigma^{\pm}_{\alpha \beta} (\varepsilon) = \frac{\hbar}{2}
\Gamma_{\mu}^{\alpha \delta} \Gamma_{\nu}^{\gamma \beta}
\int  d \Omega   \frac{1}{\Omega} g_{\mu \nu}(\Omega)
 \nonumber \\
   (  (1 - f(\varepsilon-\hbar \Omega) + n_{b}(\Omega))
G_{ \delta \gamma}^{\pm} (\varepsilon-\hbar \Omega)
\nonumber \\
+ ( f(\varepsilon + \hbar \Omega) + n_{b}(\Omega) )
G_{\delta \gamma}^{\pm} (\varepsilon+\hbar \Omega) ),
\label{self}
\end{eqnarray}
where $g_{\mu \nu}(\Omega) = \sum_{q} \langle \mu \vert q \rangle
\delta(\Omega - \Omega_q) \langle q \vert \nu \rangle$ is an element
of the vibration mode density matrix. Note again that we use Einstein
summation convention, so that summation is implied over orbital
indices $\gamma$ and $\delta$ and the phonon polarization indices
$\mu$ and $\nu.$

For simplicity, we now assume that: (i) The
bosonic coupling only affects the Cu-$d_{x^2-y^2}$ orbitals, where we 
include a diagonal self-energy of the form, $g(\Omega) = g 
\Omega^2$ when 
$\Omega \le \Omega_{d}$ and it is 0 when $\Omega >
\Omega_{d}$.  For a Debye spectrum of phonons,
$\Omega_{d}$ is the Debye cut-off frequency, and the normalization factor 
is $g=3/\Omega_{d}^{3}$; (ii) 
$\rho_{\delta \gamma} = -\frac{1}{\pi} Im[G^{+}_{\delta
\gamma}] = \rho$ is approximately a constant. This amounts to assuming
that the electronic density of states is smoothly varying within the
range of the phononic spectrum; (iii) Take $\Gamma_{\mu}^{\alpha
\delta} = \Gamma$, a constant parameter.  Using these assumptions,
the final form for the self-energy is
\begin{equation}
\Sigma^{+} =
 -\frac{A}{\pi} \left( (2z+i\pi) + \left(z^2 - 1\right)
 \ln{\left(\frac{z - 1} {z + 1} \right)}
 \right),
\label{sigmaeinstein}
\end{equation}
where $z=(\varepsilon + i \eta)/(\hbar \Omega_d)$,
$A = \frac{3 \hbar}{4 \Omega_d}  \Gamma^2 \rho,$ and $\eta$
is a convergence parameter.
Although we have derived the preceding form for coupling to a 3D Debye
spectrum of phonons, the results are not too sensitive to 
details of the spectrum, and we would expect a similar result for an 
Einstein phonon or
the magnetic resonance mode\cite{magres}.

It is interesting to consider the asymptotic forms of self-energy as 
follows. 
If $\hbar \Omega_d  \ll  \varepsilon$,
\begin{displaymath}
\Sigma(\varepsilon) \approx -A \left( \frac{2}{\pi z}
 +i \right)
\end{displaymath}
For large boson energies, i.e., $\hbar \Omega_d  \gg  \varepsilon$,
we obtain
\begin{equation}
\Sigma(\varepsilon) \approx -A \left( \frac{4}{\pi}z
 +i z^2  \right).
\label{asympt}
\end{equation}

While Eq. \eqref{self} gives a general form of phononic self-energy for 
any pair of orbitals, in the present calculations, we adopt a few 
simplifications. First, we assume only diagonal terms of self-energy to 
make the model more tractable. Second, we apply Eq. \eqref{sigmaeinstein} 
to Cu-$d_{x^2-y^2}$ orbitals using parameters $\hbar \Omega_d = 80meV$ and 
$A = 60meV.$ The former value gives the best fit to the peak-dip-hump 
structure, and the latter controls the smoothness of the spectrum.  In 
Fig. \ref{pristine}(c) we make a comparison between the more general form 
of Eq. \eqref{self} with the accurate density of states of 
Cu-$d_{x^2-y^2}$ orbitals.  For the remaining orbitals we mimic a 
Fermi-liquid type self-energy, which can be modeled with a $\Sigma'' 
\propto \varepsilon^2$ and $\Sigma' \propto \varepsilon$; here we employ 
the asymptotic form of Eq. \eqref{asympt}, choosing parameters $\hbar 
\Omega_d = 2.0eV$ (to ensure the correct asymptotic form for whole the 
energy range) and $A = 100meV.$ In this way, the need for a Kramers-Kronig 
transformation is avoided.

We can straightforwardly include in the self-energy the effect of
magnon scattering\cite{MSB} responsible for the high energy
kink\cite{HEK}. This will broaden the spectrum in the vicinity of the
VHS peaks, thereby improving agreement with experiment in
Fig. \ref{pristine}(a). It should be noted, however, that a more
accurate modeling of the self-energy will be required both for the
bosonic coupling and the Fermi-liquid term for treating the underdoped
system.

\section{Bogoliubov quasiparticles in tight-binding basis}

This appendix discusses aspects of the Bogoliubov transformation 
within a tight-binding basis. The Bogoliubov
transformation is not explicitly carried out in the present calculations
since the Green's function tensor is obtained directly from Dyson's
equation. Nevertheless, understanding the relation between the
transformation and the Green's function tensor in the tight-binding basis 
is necessary for interpreting some of our results. 
In particular, our analysis of pairing symmetry is based on the
relation between $F_{\alpha \beta}$ and $\langle c_{\alpha \uparrow}
c_{\beta\downarrow} \rangle .$

The Bogoliubov transformation \cite{bogoliubov} is conventionally
carried out in a combined basis of spin-up electrons and spin-down
holes:
\begin{equation}
\mathbf{c}_{k} = \left(
\begin{array}{c}
 c_{k \uparrow}\\
c^{\dagger}_{-k \downarrow}.
\end{array}
\right)
\end{equation}
These $c$'s diagonalize the one-particle Hamiltonian of Eq. \eqref{H1} 
via the transformations
\begin{displaymath}
c_{\alpha \uparrow} = \langle \alpha \vert k \rangle c_{k \uparrow}
\end{displaymath}
and
\begin{displaymath}
c^{\dagger}_{\alpha \downarrow} = \langle -k \vert \alpha \rangle
c^{\dagger}_{-k \downarrow} = \langle  \alpha \vert k \rangle
c^{\dagger}_{-k \downarrow},
\end{displaymath}
or in a more compact form:

\begin{equation}
\mathbf{c}_{\alpha} =
\left(
\begin{array}{c}
 c_{\alpha \uparrow}\\
c^{\dagger}_{\alpha \downarrow}.
\end{array}
\right)
=\left(
\begin{array}{cc}
 \langle  \alpha \vert k \rangle & 0\\
0 & \langle  \alpha \vert k \rangle
\end{array}
\right)
\mathbf{c}_{k}
= B_{\alpha k} \mathbf{c}_{k},
\end{equation}
with inverse $\mathbf{c}_{k} = B_{ k \alpha}
\mathbf{c}_{\alpha}.$

This change of basis diagonalizes the one-particle Hamiltonian:
\begin{displaymath}
\varepsilon_{k} = \langle k \vert \alpha \rangle
 H_{1,\alpha \beta}
\langle \beta \vert k  \rangle
\end{displaymath}
(with summation over $\alpha$ and $\beta$).
In this basis the Hamiltonian of Eq.~\eqref{hamiltonian} becomes
\begin{displaymath}
H = \varepsilon_{k} c^{\dagger}_{k \uparrow}c_{k  \uparrow} + \varepsilon_{k}
(1-c_{-k  \downarrow} c^{\dagger}_{-k \downarrow})
+ \Delta_{k} c^{\dagger}_{k \uparrow}c^{\dagger}_{-k \downarrow}
+ \Delta^{\dagger}_{k}c_{-k \downarrow} c_{k \uparrow},
\end{displaymath}
now with summation over $k.$
After shifting this by a constant energy, it assumes the simple form
\begin{displaymath}
H^{eff} = \mathbf{c}^{\dagger} \hat{H} \mathbf{c},
\end{displaymath}
where
\begin{equation}
\hat{H} = \left(
\begin{array}{cc}
 \varepsilon_{k} & \Delta_{k}\\
\Delta^{\dagger}_{k} & -\varepsilon_{k}
\end{array}
\right), 
\end{equation}
which can be diagonalized into
\begin{displaymath}
H^{eff} = \mathbf{c}^{\dagger}U^{-1} U \hat{H}U^{-1} U \mathbf{c},
\end{displaymath}
where
\begin{displaymath}
U = \left(
\begin{array}{cc}
u^{*}_{k} & v_{k}\\
-v^{*}_{k} & u_{k}
\end{array}
\right)~~\textrm{and}~~
U^{-1} = \left(
\begin{array}{cc}
u_{k} & -v_{k}\\
v^{*}_{k} & u^{*}_{k}
\end{array}.
\right)
\end{displaymath}
The coefficients are chosen in the standard way in order to obtain a 
diagonal matrix
\begin{displaymath}
 U \hat{H}U^{-1} =
\left(
\begin{array}{cc}
E_{k} & 0\\
0 & -E_{k}
\end{array}
\right),
\end{displaymath}
with $E_{k} = \sqrt{\varepsilon_k^2 + \vert \Delta_k \vert^2}.$

This Bogoliubov transformation introduces the
quasi-particle basis
\begin{displaymath}
\mathbf{a} = \left(
\begin{array}{c}
a_{k}\\
b^{\dagger}_{-k}
\end{array}
\right)
= U \mathbf{c}.
\end{displaymath}

Since we are working in the tight-binding basis, we end up with
\begin{displaymath}
 \left(
\begin{array}{c}
a_{k}\\
b^{\dagger}_{-k}
\end{array}
\right)
= \left(
\begin{array}{cc}
u^{*}_{k} \langle k \vert \alpha \rangle &
v_{k} \langle k \vert  \beta \rangle\\
-v^{*}_{k} \langle k \vert \alpha \rangle &
u_{k} \langle k \vert  \beta \rangle
\end{array}
\right)
\left(
\begin{array}{c}
 c_{\alpha \uparrow}\\
 c^{\dagger}_{\beta \downarrow}
\end{array}
\right)
\end{displaymath}
(summation over $\alpha$ and $\beta$)
or inversely
\begin{displaymath}
\left(
\begin{array}{c}
 c_{\alpha \uparrow}\\
 c^{\dagger}_{\beta \downarrow}
\end{array}
\right)
= \left(
\begin{array}{cc}
\langle \alpha \vert  k \rangle u_{k} &
-\langle \alpha \vert  k \rangle v_{k}\\
\langle   \beta \vert k \rangle v^{*}_{k} &
\langle  \beta \vert k \rangle u^{*}_{k}
\end{array}
\right)
 \left(
\begin{array}{c}
a_{k}\\
b^{\dagger}_{-k}
\end{array}
\right)
\end{displaymath}
(summation over $k$).

We are particularly interested in writing the expectation values of
electron and hole densities, $\langle c^{\dagger}_{\alpha \sigma}c_{\beta
\sigma}\rangle,$ and $\langle c_{\alpha \sigma}c^{\dagger}_{\beta \sigma}
\rangle,$ and pairing amplitudes $\langle c^{\dagger}_{\alpha
\uparrow}c^{\dagger}_{\beta \downarrow} \rangle,$ and $\langle
c_{\beta \downarrow} c_{\alpha \uparrow} \rangle$ in
terms of the Green's function tensor. For this purpose, we start with a 
$2\times 2$ tensor
\begin{equation}
 \langle \mathbf{c}_{\alpha} \mathbf{c}^{\dagger}_{\beta} \rangle =
 \langle B_{\alpha k} \mathbf{c}_{k} \mathbf{c}^{\dagger}_{k}B_{k
 \beta} \rangle = \langle B_{\alpha k}
 U^{-1}\mathbf{a}\mathbf{a}^{\dagger} U B_{k \beta}\rangle.
\label{Ntensor}
 \end{equation}
Using the fact that $\langle a_k a^{\dagger}_k \rangle = 1-f(E_k)$
 and $\langle b^{\dagger}_k b_k \rangle = f(E_k)$, we evaluate each
 element of the tensor $ \langle \mathbf{c}_{\alpha}
 \mathbf{c}^{\dagger}_{\beta} \rangle $ separately as follows:\\
\indent
(1) The number density
\begin{displaymath}
\langle c^{\dagger}_{\alpha \uparrow}c_{\beta \uparrow}
\rangle = \langle \beta \vert k \rangle \left( \vert u \vert^2 f(E_k)
+ \vert v \vert^2 (1-f(E_k)) \right) \langle k \vert \alpha \rangle
\end{displaymath}
Now we use a trick following Ref. \onlinecite{Horsfield} where
\begin{displaymath}
 \langle \beta \vert k \rangle \vert u \vert^2 f(E_k) \langle k \vert \alpha
 \rangle = \int d\varepsilon f(\varepsilon) \langle \beta \vert k
 \rangle  u  \delta(\varepsilon - E_{k}) u^{*} \langle k \vert \alpha
 \rangle
\end{displaymath}
and
 $$\delta(\varepsilon - E_{k}) \approx
-\frac{1}{\pi}Im(\frac{1}{\varepsilon -E_{k} +i \eta}).$$ Hence
\begin{displaymath}
 \langle \beta \vert k \rangle \vert u \vert^2 f(E_k) \langle k \vert
 \alpha \rangle = \int d\varepsilon f(\varepsilon) \rho^{e}_{\beta
 \alpha}(\varepsilon),
\end{displaymath}
where
\begin{displaymath}
 \rho^{e}_{\beta \alpha}(\varepsilon) = -\frac{1}{\pi}Im(G^{+}_{e,
 \beta \alpha}(\varepsilon)),
\end{displaymath}
where $G^{+}_{e, \beta \alpha}$ refers to the electron part of the
Green's function,
\begin{displaymath}
  G^{+}_{e, \alpha \beta}(\varepsilon) =
\frac{\langle \alpha, e \vert k \rangle
\langle k \vert e, \beta \rangle}{\varepsilon - E_k + i \eta}
  = \frac{\langle \alpha \vert k \rangle   \vert u_{k} \vert^{2}
\langle k  \vert \beta \rangle} {\varepsilon - E_k + i \eta}.
\end{displaymath}

It is straightforward to show that
\begin{displaymath}
\langle c^{\dagger}_{\alpha \uparrow}c_{\beta \uparrow} \rangle = \int
d\varepsilon [f(\varepsilon) \rho^{e}_{\beta \alpha}(\varepsilon) +
(1-f(\varepsilon)) \rho^{h}_{\beta \alpha}(\varepsilon)],
\end{displaymath}
where $\rho^{h}_{\alpha \beta}$ is the hole density matrix.  The first
part of the integral, in fact, gives the number of electrons with a
chosen spin. The latter part gives the same result as the former
since the Bogoliubov transformation reflects the electron bands to hole
bands with respect to the Fermi energy;\\
\indent
(2) The pairing amplitude
\begin{displaymath}
\langle c_{\alpha \uparrow}c_{\beta \downarrow}
\rangle = \langle \alpha \vert k \rangle \left( u  (f(E_k) -
 (1-f(E_k)) v \right) \langle k \vert \beta \rangle .
\end{displaymath}
Using the trick of Ref. \onlinecite{Horsfield} again gives us the formula
\begin{equation}
\langle c_{\alpha \uparrow}c_{\beta \downarrow} \rangle = -\int
d\varepsilon (1-2f(\varepsilon)) \rho^{eh}_{\alpha \beta}(\varepsilon),
\label{ccF}
\end{equation}
where
\begin{displaymath}
 \rho^{eh}_{\alpha \beta}(\varepsilon) = -\frac{1}{\pi}Im(F^{+}_{\alpha
 \beta}(\varepsilon)),
\end{displaymath}
and
\begin{displaymath}
  F^{+}_{\alpha \beta}(\varepsilon) =
\frac{\langle \alpha, e \vert k \rangle
\langle k \vert h, \beta \rangle}{\varepsilon - E_k + i \eta}
  = \frac{\langle \alpha \vert k \rangle  u_{k} v_{k}
\langle k   \vert \beta \rangle}{\varepsilon - E_k + i \eta}
\end{displaymath}

 In the same manner, one can see that
\begin{equation}
\langle c^{\dagger}_{\alpha \uparrow}c^{\dagger}_{\beta \downarrow}
\rangle = -\int d\varepsilon (1-2f(\varepsilon)) \rho^{eh\dagger}_{\beta
\alpha}(\varepsilon),
\label{ccF1}
\end{equation}
where
\begin{displaymath}
 \rho^{eh\dagger}_{\beta \alpha}(\varepsilon) =
 -\frac{1}{\pi}Im((F^{+})^{\dagger}_{\beta \alpha}(\varepsilon)),
\end{displaymath}

Equations \eqref{ccF} and \eqref{ccF1} also reveal how the anomalous part 
of the Green's function tensor is related to the pairing amplitude 
$\langle c_{\alpha \uparrow}c_{\beta \downarrow} \rangle$ in a 
tight-binding basis, or equivalently how the anomalous part of the current 
is related to the making and breaking of Cooper pairs. In particular, 
symmetry properties of $F_{\alpha \beta}$ are seen to be related directly 
to those of $\langle c_{\alpha \uparrow}c_{\beta \downarrow} \rangle$.



\begin{thebibliography}{99}
  
\bibitem{Fischer} \O. Fischer, M. Kugler, I. Maggio-Aprile, and Chr.
  Berthod, and Chr. Renner, {\it Rev. Mod. Phys.} {\bf 79}, 353
  (2007).

\bibitem{McElroy} K. McElroy,
 Jinho Lee, J.A. Slezak, D.-H. Lee, H. Eisaki, S. Uchida, and J.C. Davis,
{\it Science} {\bf 309}, 1048 (2005).

\bibitem{Hudson} E.W. Hudson,
 K.M. Lang, V. Madhave, S.H. Pan, H. Eisaki, S. Uchida, and J.C. Davis,
{\it Nature} {\bf 411}, 920 (2001).

\bibitem{Pan} S.H. Pan, E.W. Hudson, K.M. Lang, H. Eisaki, S. Uchida,
and J.C. Davis, {\it Nature} {\bf 403}, 746(2000).

\bibitem{Yazdani}A.N. Pasupathy, A. Pushp, K.K. Gomes, C.V. Parker,
J. Wen, Z. Xu, G. Gu, S. Ono, Y. Ando, and A. Yazdani, {\it Science}
{\bf 320}, 196 (2008).

\bibitem{Balatsky1}A.V. Balatsky,
 , A. V., Vekhter, I., and Zhu, J.-X.,
{\it Rev. Mod. Phys.} {\bf 78}, 373 (2006).

 \bibitem{Kohsaka}Y. Kohsaka, C. Taylor, K. Fujita, A. Schmidt, C. Lupien,
T. Hanaguri, M. Azuma, M. Takano, H. Eisaki, H. Takagi, S. Uchida,
 and J.C.  Davis, {\it Science} {\bf 315}, 1380 (2007).


 \bibitem{Balatsky} I. Martin, A.V. Balatsky, and J. Zaanen,
{\it Phys. Rev. Lett.} {\bf 88}, 097003 (2002).


 \bibitem{hoogenboom} B.W.~Hoogenboom, C.~Berthod, M.~Peter, \O.~ Fischer,
 and A.A.~Kordyuk, {\it Phys. Rev. B} {\bf 67}, 224502(2003).

\bibitem{NLMB} J.A. Nieminen, H. Lin, R.S. Markiewicz, and A. Bansil,
 {\it Phys. Rev. Lett.} {\bf 102},037001 (2009).

 \bibitem{Todorov} T.N.~Todorov, G.A.D.~Briggs and A.P.~Sutton,
{\it J.Phys.: Condens. Matter} {\bf 5}, 2389 (1993).

 \bibitem{Pendry} J.B.~Pendry, A.B.~Pr\^etre and B.C.H.~Krutzen,
{\it J.Phys.: Condens. Matter} {\bf 3}, 4313 (1991).

\bibitem{Tersoff} J. Tersoff and D.R. Hamann,
{\it Phys. Rev. B} {\bf 31}, 805 (1985).

\bibitem{footnote1} Note that the tunneling signal decays
exponentially with layer distance from the tip, and therefore, we expect
the results presented in this article to be essentially the same as
for a semi-infinite solid.

 \bibitem{Bellini} V.~Bellini,
 F.~Manghi, T.~Thonhauser, and C.~Ambrosch-Draxl,
{\it Phys. Rev. B} {\bf 69}, 184508(2004).



\bibitem{Slater}  J.C. Slater and  G.F. Koster,
{\it Phys. Rev.} {\bf 94}, 1498 (1954).



\bibitem{Harrison} W.A. Harrison, {\it Electronic Structure and
    Properties of Solids.} Dover, New York (1980).

 \bibitem{Shi} L.~Shi and D.~A.~Papaconstantopoulos,
 {\it Phys. Rev. B} {\bf 70}, 205101 (2004).

%

\bibitem{bansil99} A. Bansil and M. Lindroos, {\it Phys. Rev. Lett.} 83, 5154(1999).
  
\bibitem{lindroos02} M. Lindroos, S. Sahrakorpi and A. Bansil, {\it Phys.
  Rev. B} {\bf 65}, 054514(2002)
  
\bibitem{bansil05} A. Bansil, M. Lindroos, S. Sahrakorpi, and R.S.
  Markiewicz, {\it Phys.  Rev. B} {\bf 71}, 012503(2005).
  
\bibitem{markiewicz05} R.S. Markiewicz, S. Sahrakorpi, M. Lindroos,
  Hsin Lin, and A. Bansil, {\it Phys. Rev. B} {\bf 72}, 054519(2005).
  
\bibitem{asensio03} M.C. Asensio, J. Avila, L. Roca, A. Tejeda, G. D.
  Gu, M. Lindroos, R. S. Markiewicz, and A. Bansil,
{\it Phys. Rev. B} {\bf 67}, 014519(2003).
  
\bibitem{bansil98} A. Bansil and M. Lindroos, {\it Journal of Physics
    and Chemistry of Solids} {\bf 59}, 1879(1998).

\bibitem{HL} H. Lin, S. Sahrakorpi, R.S. Markiewicz, and A. Bansil,
{\it Phys. Rev. Lett.} {\bf 96}, 097001 (2006).


\bibitem{Gomes} K. K. Gomes, A. N. Pasupathy, A.
  Pushp, S. Ono, Y. Ando, and A. Yazdani, Nature {\bf 447}, 569-572(2007).

\bibitem{Kaminski} A. Kaminski et al., {\it Phys. Rev. B} {\bf 73},
174511(2006).


\bibitem{Nieminen} J.A. Nieminen and S. Paavilainen,
{\it Phys. Rev. B} {\bf 60}, 2921 (1999).


\bibitem{footfourier} In practice, we calculate the Green's function
  $G^{e}_{\mathbf{k} \alpha \beta}$ for each k-point separately to
  produce the site-dependent Green's function by inverse
  Fourier-transformation: $G^{e}_{i\alpha, j\beta} = \frac{1}{N_k}
  \sum_{k}G^{e}_{\mathbf{k} \alpha \beta} \exp{(-i\mathbf{k} \cdot
  \mathbf{R}_{ij})}.$ Here the shorthand notation, $G_{e,\alpha \beta}
  = G^{e}_{i\alpha, j\beta}$, is used in that indices $\alpha$ and
  $\beta$ implicitly contain the simulation cell index. Note also,
  that the inverse transformation must not be done until solving the
  whole Green's function tensor.

  
\bibitem{Flatte} J.-M. Tang and M. E. Flatt\'e, {\it Phys. Rev. B}
  {\bf 66}, 060504(R) (2002); J.-M. Tang and M. E. Flatt\'e, {\it
    Phys. Rev. B} {\bf 70}, 140510(R) (2004).

\bibitem{Fetter}  A.L. Fetter and  J.D. Walecka,
{\it Quantum Theory of Many-Particle Systems.} Dover (2003).

\bibitem{Zhang} F.C. Zhang and T.M. Rice, {\it Phys. Rev. B} {\bf 37},
  3759(1988).

\bibitem{onsitep}
The d-wave and ZRS symmetries are
not uniquely determined by the present choice of pairing. For instance, we
could choose an onsite pairing at the oxygen $p_{x/y}$ orbitals
with $\Delta_{xx} = -\Delta_{yy}.$
In that case, Eq. \eqref{fdelta}
could be written in the $x$-direction as, 
$$F_{ \alpha \beta} = G_{e,\alpha x} \Delta_{xx} G^{0}_{h,x\beta},$$
with a corresponding expression in the
$y$-direction. If $\alpha = d$ and $\beta = d\pm x$, the sign of
$G_{e,\alpha x}G^{0}_{h,x \beta}$ is totally determined by the
product of the lobes of the $d$ orbitals of the neighboring Cu atoms.
This is, however, invariant under rotation by $\pi/2$, and thus, 
$F_{d,d\pm x /d\pm y}$ follows the symmetry of $\Delta_{xx / yy}.$


\bibitem{Meir}   Y. Meir and N.S. Wingreen,
{\it Phys. Rev. Lett.} {\bf 68}, 2512 (1992).


\bibitem{Fisher} H. Ness and A.J. Fisher,
{\it Phys. Rev. B} {\bf 56}, 12469 (1997).


 \bibitem{Frederiksen} T.~Frederiksen, M.~Paulsson, M.~Brandbyge, and
 A.-P.~Jauho,
{\it Phys. Rev. B} {\bf 75}, 205413 (2007).


\bibitem{footdecomposition} In the present study, the decomposition of
Eq. \eqref{conductance} into tunneling channels has been significantly
elaborated beyond our recent work in Ref. \onlinecite{NLMB}. The most
significant improvement is the explicit formulation of anomalous
tunneling channels. Furthermore, we can obtain 
the total contribution of any chosen orbital over the whole infinite 
slab via the relation
$  \sum_{i=0}^{\infty} G^{+}_{0 c i \alpha}\Sigma{''}_{i
  \alpha}G^{-}_{i \alpha 0 c'} = \frac{1}{N_k} \sum_{\mathbf{k}}
  G^{+}_{\mathbf{k} c \alpha}\Sigma{''}_{\alpha}G^{-}_{\mathbf{k}
  \alpha c'}$
to regular and anomalous terms of Eq. \eqref{singlespec}. Note that
the simulation cell indices $0$ and $i$ are explicitly written in the
left hand side.


\bibitem{Magoga} M.~Magoga and C.~Joachim, {\it Phys. Rev. B}
{\bf 59}, 16011 (1999).


\bibitem{Sautet}  P. Sautet, {\it Surf. Sci.} {\bf 374}, 374 (1997).

\bibitem{Niemi} E. Niemi and J. Nieminen, {\it Chem. Phys. Lett.} {\bf 397},
  200 (2004).

\bibitem{footnote2} The distinct VHS peaks in the computed spectrum
are expected to be broadened to yield a smooth hump much like the
experimental spectrum due to self-energy corrections resulting from
magnetic response of the electron gas in the -400 meV range.  These
self-energy corrections are not included in the present calculations.

\bibitem{MSB} R.S. Markiewicz, S. Sahrakorpi, and A. Bansil, {\it Phys. Rev. B}
{\bf 76}, 174514 (2007).


\bibitem{footkpoints} Note that in solving the Dyson's equation, the
  initial Green's function for each k-point is a diagonal matrix of
  complex Lorentzians. An infinite number of k-points would be
  required for a final Green's function without any artificial
  ``shoulders'', although the imaginary part of the self-energy acts
  to smooth away the unwanted structures.

  
\bibitem{kinks} A. Lanzara et al., {\it Nature} (London) {\bf 412},
  510 (2001); X. J. Zhou et al., {\it Phys. Rev. Lett.} {\bf 95},
  117001 (2005).  A. Kaminski et al., {\it Phys. Rev. Lett.} {\bf 86},
  1070 (2001); P. D. Johnson et al., ibid. {\bf 87}, 177007 (2001); S. V.
  Borisenko et al., ibid. {\bf 90}, 207001 (2003); A. D. Gromko et al.,
  {\it Phys. Rev. B} {\bf 68}, 174520 (2003).


 \bibitem{VHS} G. Levy de Castro, Chr. Berthod, A. Piriou,
 E. Giannini, and \O. Fischer,
  {\it Phys. Rev. Lett.} {\bf 101}, 267004 (2008).


\bibitem{Rieder} L. Bartels, G. Meyer and K.-H. Rieder. {\it Appl. Phys.
  Lett.} 71, 213 (1997).

\bibitem{Niemi1} J. Nieminen, E. Niemi, K.-H. Rieder,
  {\it Surface Science} {\bf 552}, L47-L52 (2004).



\bibitem{THTPfoot} Although TP and TH are formulated in terms of LDOS,
  both approaches can, in principle, be decomposed into a spectral 
function
  form since the density matrix can be decomposed in this way 
regardless of the basis set employed. 


\bibitem{ShenRMP}A.P. Shen, {\it Rev. Mod. Phys.} {\bf 4}, 382 (1971).




\bibitem{paulsson-05} M. Paulsson, T. Frederiksen, and M.  Brandbyge,
  {\it Phys. Rev. B} {\bf 72}, 201101(R) (2005).
  
\bibitem{magres}Z.-X. Shen and J.R. Schrieffer, {\it Phys. Rev. Lett.}
  {\bf 78}, 1771 (1997); M.R. Norman and H. Ding, {\it Physical Review
    B} {\bf 57}, R11089 (1998); S. LaShell, E. Jensen, and T.
  Balasubramanian, {\it Phys. Rev. B} {\bf 61}, 2371 (2000)
  
\bibitem{HEK} F. Ronning, K.M. Shen, N.P. Armitage, A. Damascelli,
  D.H.  Lu, Z.-X. Shen, L.L. Miller, and C. Kim, {\it Phys. Rev. B}
  {\bf 71}, 094518 (2005); J. Graf, G.-H. Gweon, K. McElroy, S.Y.
  Zhou, C. Jozwiak, E. Rotenberg, A. Bill, T. Sasagawa, H. Eisaki, S.
  Uchida, H. Takagi, D.-H. Lee, and A. Lanzara, {\it Phys. Rev. Lett.}
  {\bf 98}, 067004 (2007).

\bibitem{bogoliubov} M. Tinkham, {\it Introduction to
    Superconductivity.} McGraw-Hill International Editions (1996).

\bibitem{Horsfield} A.P. Horsfield,
A.M. Bratkovsky, M. Fearn, D.G. Pettifor, and M. Aoki,
{\it Phys. Rev. B} {\bf 53}, 12694(1996).



\end{thebibliography}
\end{document}